\documentclass[twocolumn,showpacs,preprintnumbers,amsmath,amssymb]{revtex4}

\usepackage{amsmath,amssymb,graphicx}
\usepackage{graphicx}
\usepackage{dcolumn}
\usepackage{bm}
\usepackage{epsfig}
\usepackage{here}
\usepackage{float}
\usepackage{amsmath}
\usepackage[usenames]{color}

\newcommand{\bea}{\begin{eqnarray}}
\newcommand{\eea}{\end{eqnarray}}

\begin{document}
\draft

\title{Fully nonlinear cosmological perturbations of multi-component fluid and field systems}
\author{Jai-chan Hwang$^{1}$, Hyerim Noh$^{2}$, Chan-Gyung Park$^{3}$}
\address{${}^{1}$Department of Astronomy and Atmospheric Sciences,
         Kyungpook National University, Daegu, 702-701, Korea\\
         ${}^{2}$Korea Astronomy and Space Science Institute, Daejeon, 305-348, Korea\\
         ${}^{3}$Division of Science Education and Institute of Fusion Science, Chonbuk National University, Jeonju 561-756, Republic of Korea}

\date{\today}

\begin{abstract}

We present fully nonlinear and exact cosmological perturbation equations in the presence of multiple components of fluids and minimally coupled scalar fields. We ignore the tensor-type perturbation. The equations are presented without taking the temporal gauge condition in the Friedmann background with general curvature and the cosmological constant. For each fluid component we ignore the anisotropic stress. The multiple component nature, however, introduces the anisotropic stress in the collective fluid quantities. We prove the Newtonian limit of multiple fluids in the zero-shear gauge and the uniform-expansion gauge conditions, present the Newtonian hydrodynamic equations in the presence of general relativistic pressure in the zero-shear gauge, and present the fully nonlinear equations and the third-order perturbation equations of the nonrelativistic pressure fluids in the CDM-comoving gauge.

\end{abstract}

\maketitle

\section{Introduction}
                                             \label{sec:Introduction}

Cosmological perturbation theory is an important tool in mediating cosmological models with observations. The spatially homogeneous and isotropic world model based on the cosmological constant modified Einstein's gravity with small perturbations (Friedmann 1922, Lifshitz 1946) enjoyed unexpected and unprecedented success in explaining most of the cosmologically relevant observations. Minor or major (depending on one's view) shortcomings of this $\Lambda$CDM (cosmological constant and cold dark matter) model are, besides the small-scale problem of the CDM model, that substantial amounts of dark energy and dark matter are needed to maintain the model confronted by observations. The cosmological constant is a minimal form of the dark energy (often modelled using a scalar field), but this should be considered as a pure theoretical modification of Einstein's gravity introduced by Einstein himself (Einstein 1917). Besides the dark atoms (invisible to light detectors), substantial amount of the dark matter is also currently in a state of unidentified character.

With the help of such minor/major unknowns the concordance comological model (the $\Lambda$CDM Friedmann world model with small perturbations) is agreed among researchers to be quite successful in delineating the structure and evolution of the large-scale observable universe. As the scale becomes smaller and the time goes over into later stage, the nonlinear processes cannot be ignored, but these processes are often treated in the Newtonian context using either nonlinear perturbations or numerical simulations (Vishniac 1983, Blumenthal et al 1984, Bernardeau et al 2002, L'Huillier, Park \& Kim 2014). Nonlinear perturbation theory in Einstein's gravity also has been developed in parallel with increasing observational precisions (Bartolo et al 2004, Noh \& Hwang 2004, Malik \& Wands 2009).

Fully nonlinear formulation of cosmological perturbation theory in the context of Einstein's gravity was recently introduced (Hwang \& Noh 2013a, Noh 2014). The formulation ignores spatially transverse-tracefree (tensor-type) perturbation, but have not taken the slicing (temporal gauge) condition.  Besides simple derivation of the perturbation equations to any nonlinear order, the formulation was used to show the Newtonian limit, the first-order post-Newtonian approximation and the Newtonian hydrodynamic equations with general relativistic (gravitating) pressure (Hwang \& Noh 2013b, 2013c, Noh \& Hwang 2013). The formulation considered a fluid (without anisotropic stress) or a scalar field system.

Here we extend the fully nonlinear perturbation formulation to include the multiple component fluids (still without anisotropic stress of individual fluid) and multiple component minimally coupled scalar fields.

We set $c \equiv 1 \equiv \hbar$ for the scalar fields.

\section{Fully nonlinear perturbations of a fluid}
                                                                                \label{sec:equations}

Our metric convention in the perturbation theory is (Bardeen 1988, Hwang \& Noh 2013a)
\bea
   & & ds^2 = - a^2 \left( 1 + 2 \alpha \right) (d x^0)^2
       - 2 a \chi_{i} d x^0 d x^i
   \nonumber \\
   & & \qquad
       + a^2 \left( 1 + 2 \varphi \right) \gamma_{ij} d x^i d x^j,
   \label{metric-PT}
\eea
where the spatial index of $\chi_i$ is raised and lowered by $\gamma_{ij}$ as the metric; $\gamma_{ij}$ is the comoving part of the three-space metric of the Robertson-Walker metric. Here we {\it assume} $a$ to be a function of time only, and $\alpha$, $\varphi$ and $\chi_i$ are arbitrary functions of spacetime. The spatial part of the metric is simple because we already have taken the spatial gauge condition without losing any generality (Bardeen 1988, Hwang \& Noh 2013a), and have ignored the tensor-type perturbation.
The energy-momentum tensor of a fluid is (Ehlers 1961, Ellis 1971, Ellis 1973)
\bea
   & & \widetilde T_{ab} = \widetilde \mu \widetilde u_a \widetilde u_b
       + \widetilde p \widetilde h_{ab}
       + \widetilde q_a \widetilde u_b
       + \widetilde q_b \widetilde u_a
       + \widetilde \pi_{ab},
   \label{Tab}
\eea
where $\widetilde u_a$ is the normalized four-vector with $\widetilde u^a \widetilde u_a \equiv -1$ and $\widetilde h_{ab} \equiv \widetilde g_{ab} + \widetilde u_a \widetilde u_b$ is the projection tensor. Tildes indicate covariant quantities; $\widetilde \mu$, $\widetilde p$, $\widetilde q_a$ and $\widetilde \pi_{ab}$ are the covariant energy density, pressure, flux vector and anisotropic stress tensor, respectively. The fluid quantities in equation (\ref{Tab}) can be read as
\bea
   & & \widetilde \mu = \widetilde T_{ab} \widetilde u^a \widetilde u^b, \quad
       \widetilde p = {1 \over 3} \widetilde T_{ab} \widetilde h^{ab},
   \nonumber \\
   & &
       \widetilde q_a = - \widetilde T_{cd} \widetilde u^c
       \widetilde h^d_a, \quad
       \widetilde \pi_{ab} = \widetilde T_{cd} \widetilde h^c_a \widetilde h^d_b - \widetilde p \widetilde h_{ab}.
   \label{Tab-decomposition}
\eea
We will take the energy-frame condition $\widetilde q_a \equiv 0$ without losing any generality; in this case $\widetilde u_a$ becomes the fluid four-vector.
In the perturbation theory we may introduce (Hwang \& Noh 2013a)
\bea
   & & \widetilde \mu \equiv \mu + \delta \mu, \quad
       \widetilde p \equiv p + \delta p, \quad
       \widetilde \pi_{ij} \equiv a^2 \Pi_{ij},
   \nonumber \\
   & &
       \widetilde u_i \equiv a \widehat \gamma {\widehat v_i \over c}, \quad
       \widehat \gamma \equiv {1 \over \sqrt{ 1 - {\widehat v^k \widehat v_k
       \over c^2 (1 + 2 \varphi)}}},
   \label{fluid-PT}
\eea
where $\mu$ and $p$ are the background energy density and pressure, respectively; spatial indices of $\widehat v_i$ and $\Pi_{ij}$ are raised and lowered by $\gamma_{ij}$ as the metric; the perturbed fluid quantities $\delta \mu$, $\delta p$, $\widehat v_i$ and $\Pi_{ij}$ are functions of space-time with arbitrary amplitudes.
In the following we will keep $\widetilde \mu$ and $\widetilde p$ without decomposition. We can decompose $\chi_i$, $v_i$ and $\Pi_{ij}$ to scalar- and vector-type perturbations as (Bardeen 1980, Bardeen 1988, Noh \& Hwang 2004, Hwang \& Noh 2013a)
\bea
   & & \chi_i \equiv \chi_{,i} + \chi^{(v)}_i, \quad
       \widehat v_i \equiv - \widehat v_{,i} + \widehat v^{(v)}_i,
   \nonumber \\
   & &
       \Pi_{ij} \equiv {1 \over a^2} \left(
       \Pi_{,i|j} - {1 \over 3} \gamma_{ij} \Delta \Pi \right)
       + {1 \over a} \Pi^{(v)}_{(i|j)}
       + \Pi^{(t)}_{ij},
   \label{s-v}
\eea
with the vector-type perturbations satisfying $\chi^{(v)i}_{\;\;\;\;\;\;|i} = \widehat v^{(v)i}_{\;\;\;\;\;\;|i} = \Pi^{(v)i}_{\;\;\;\;\;\;|i} \equiv 0$ and $\Pi^{(t)i}_{\;\;\;\;i} \equiv 0 \equiv \Pi^{(t)j}_{\;\;\;\;i|j}$; a vertical bar indicates a covariant derivative based on the metric $\gamma_{ij}$. To the nonlinear order our scalar- and vector-type perturbations are coupled in the equation level.
In Hwang \& Noh (2013a) we presented nonlinear perturbation equations for vanishing anisotropic stress in a flat background. Here, we consider background world model with general curvature (Noh 2014). The exact and fully nonlinear perturbation equations, without taking the temporal gauge (slicing or hypersurface) condition are the following.

\begin{widetext}
\noindent
Definition of $\kappa$: \bea
   & & \kappa
       \equiv
       3 {\dot a \over a} \left( 1 - {1 \over {\cal N}} \right)
       - {1 \over {\cal N} (1 + 2 \varphi)}
       \left[ 3 \dot \varphi
       + {c \over a^2} \left( \chi^k_{\;\;|k}
       + {\chi^{k} \varphi_{,k} \over 1 + 2 \varphi} \right)
       \right].
   \label{eq1}
\eea
ADM energy constraint:
\bea
   & & - {3 \over 2} \left( {\dot a^2 \over a^2}
       - {8 \pi G \over 3 c^2} \widetilde \mu
       + {\overline K c^2 \over a^2 (1 + 2 \varphi)}
       - {\Lambda c^2 \over 3} \right)
       + {\dot a \over a} \kappa
       + {c^2 \Delta \varphi \over a^2 (1 + 2 \varphi)^2}
   \nonumber \\
   & & \qquad
       = {1 \over 6} \kappa^2
       - {4 \pi G \over c^2} \left( \widetilde \mu + \widetilde p \right)
       \left( \widehat \gamma^2 - 1 \right)
       + {3 \over 2} {c^2 \varphi^{|i} \varphi_{,i} \over a^2 (1 + 2 \varphi)^3}
       - {c^2 \over 4} \overline{K}^i_j \overline{K}^j_i
       - {1 \over (1 + 2 \varphi)^2}
       {4 \pi G \over c^4} \Pi_{ij}
       \widehat v^i \widehat v^j.
   \label{eq2}
\eea
ADM momentum constraint:
\bea
   & & {2 \over 3} \kappa_{,i}
       + {c \over a^2 {\cal N} ( 1 + 2 \varphi )}
       \left[ {1 \over 2} \left( \Delta \chi_i
       + \chi^k_{\;\;|ik} \right)
       - {1 \over 3} \chi^k_{\;\;|ki} \right]
       + {8 \pi G \over c^4} \left( \widetilde \mu + \widetilde p \right)
       a \widehat \gamma^2 \widehat v_i
   \nonumber \\
   & & \qquad
       =
       {c \over a^2 {\cal N} ( 1 + 2 \varphi)}
       \Bigg\{
       \left( {{\cal N}_{,j} \over {\cal N}}
       - {\varphi_{,j} \over 1 + 2 \varphi} \right)
       \left[ {1 \over 2} \left( \chi^{j}_{\;\;|i} + \chi_i^{\;|j} \right)
       - {1 \over 3} \delta^j_i \chi^k_{\;\;|k} \right]
       - {\varphi^{,j} \over (1 + 2 \varphi)^2}
       \left( \chi_{i} \varphi_{,j}
       + {1 \over 3} \chi_{j} \varphi_{,i} \right)
   \nonumber \\
   & & \qquad
       + {{\cal N} \over 1 + 2 \varphi} \nabla_j
       \left[ {1 \over {\cal N}} \left(
       \chi^{j} \varphi_{,i}
       + \chi_{i} \varphi^{|j}
       - {2 \over 3} \delta^j_i \chi^{k} \varphi_{,k} \right) \right]
       \Bigg\}
       - {a \over 1 + 2 \varphi}
       {8 \pi G \over c^4} \Pi_{ij}
       \widehat v^j.
   \label{eq3}
\eea
Trace of ADM propagation:
\bea
   & & - 3 \left[ {1 \over {\cal N}}
       \left( {\dot a \over a} \right)^{\displaystyle\cdot}
       + {\dot a^2 \over a^2}
        + {4 \pi G \over 3 c^2} \left( \widetilde \mu + 3 \widetilde p \right)
       - {\Lambda c^2 \over 3} \right]
       + {1 \over {\cal N}} \dot \kappa
       + 2 {\dot a \over a} \kappa
       + {c^2 \Delta {\cal N} \over a^2 {\cal N} (1 + 2 \varphi)}
       = {1 \over 3} \kappa^2
       + {8 \pi G \over c^2} \left( \widetilde \mu + \widetilde p \right)
       \left( \widehat \gamma^2 - 1 \right)
   \nonumber \\
   & & \qquad
       - {c \over a^2 {\cal N} (1 + 2 \varphi)} \left(
       \chi^{i} \kappa_{,i}
       + c {\varphi^{|i} {\cal N}_{,i} \over 1 + 2 \varphi} \right)
       + c^2 \overline{K}^i_j \overline{K}^j_i
       + {1 \over 1 + 2 \varphi}
       {4 \pi G \over c^2} \left( \Pi^i_i
       + {1 \over 1 + 2 \varphi} \Pi_{ij}
       {\widehat v^i \widehat v^j \over c^2} \right).
   \label{eq4}
\eea
Tracefree ADM propagation:
\bea
   & & \left( {1 \over {\cal N}} {\partial \over \partial t}
       + 3 {\dot a \over a}
       - \kappa
       + {c \chi^{k} \over a^2 {\cal N} (1 + 2 \varphi)} \nabla_k \right)
       \Bigg\{ {c \over a^2 {\cal N} (1 + 2 \varphi)}
   \nonumber \\
   & & \qquad
       \times
       \left[
       {1 \over 2} \left( \chi^i_{\;\;|j} + \chi_j^{\;|i} \right)
       - {1 \over 3} \delta^i_j \chi^k_{\;\;|k}
       - {1 \over 1 + 2 \varphi} \left( \chi^{i} \varphi_{,j}
       + \chi_{j} \varphi^{|i}
       - {2 \over 3} \delta^i_j \chi^{k} \varphi_{,k} \right)
       \right] \Bigg\}
   \nonumber \\
   & & \qquad
       - {c^2 \over a^2 ( 1 + 2 \varphi)}
       \left[ {1 \over 1 + 2 \varphi}
       \left( \nabla^i \nabla_j - {1 \over 3} \delta^i_j \Delta \right) \varphi
       + {1 \over {\cal N}}
       \left( \nabla^i \nabla_j - {1 \over 3} \delta^i_j \Delta \right) {\cal N} \right]
   \nonumber \\
   & & \qquad
       =
       {8 \pi G \over c^2} \left( \widetilde \mu + \widetilde p \right)
       \left[ {\widehat \gamma^2 \widehat v^i \widehat v_j \over c^2 (1 + 2 \varphi)}
       - {1 \over 3} \delta^i_j \left( \widehat \gamma^2 - 1 \right)
       \right]
       + {c^2 \over a^4 {\cal N}^2 (1 + 2 \varphi)^2}
       \Bigg[
       {1 \over 2} \left( \chi^{i|k} \chi_{j|k}
       - \chi_{k|j} \chi^{k|i} \right)
   \nonumber \\
   & & \qquad
       + {1 \over 1 + 2 \varphi} \left(
       \chi^{k|i} \chi_k \varphi_{,j}
       - \chi^{i|k} \chi_j \varphi_{,k}
       + \chi_{k|j} \chi^k \varphi^{|i}
       - \chi_{j|k} \chi^i \varphi^{|k} \right)
       + {2 \over (1 + 2 \varphi)^2} \left(
       \chi^{i} \chi_{j} \varphi^{|k} \varphi_{,k}
       - \chi^{k} \chi_{k} \varphi^{|i} \varphi_{,j} \right) \Bigg]
   \nonumber \\
   & & \qquad
       - {c^2 \over a^2 (1 + 2 \varphi)^2}
       \Bigg[ {3 \over 1 + 2 \varphi}
       \left( \varphi^{|i} \varphi_{,j}
       - {1 \over 3} \delta^i_j \varphi^{|k} \varphi_{,k} \right)
       + {1 \over {\cal N}} \left(
       \varphi^{|i} {\cal N}_{,j}
       + \varphi_{,j} {\cal N}^{|i}
       - {2 \over 3} \delta^i_j \varphi^{|k} {\cal N}_{,k} \right) \Bigg]
   \nonumber \\
   & & \qquad
       + {1 \over 1 + 2 \varphi}
       {8 \pi G \over c^2} \left( \Pi^i_j
       - {1 \over 3} \delta^i_j \Pi^k_k \right).
   \label{eq5}
\eea
We have \bea
   & & {\cal N} \equiv \sqrt{ 1 + 2 \alpha
       + {\chi^k \chi_k \over a^2 ( 1 + 2 \varphi )}}, \quad
       \overline{K}^i_j \overline{K}^j_i
       = {1 \over a^4 {\cal N}^2 (1 + 2 \varphi)^2}
       \Bigg\{
       {1 \over 2} \chi^{i|j} \left( \chi_{i|j} + \chi_{j|i} \right)
       - {1 \over 3} \chi^i_{\;\;|i} \chi^j_{\;\;|j}
   \nonumber \\
   & & - {4 \over 1 + 2 \varphi} \left[
       {1 \over 2} \chi^i \varphi^{|j} \left(
       \chi_{i|j} + \chi_{j|i} \right)
       - {1 \over 3} \chi^i_{\;\;|i} \chi^j \varphi_{,j} \right]
       + {2 \over (1 + 2 \varphi)^2} \left(
       \chi^{i} \chi_{i} \varphi^{|j} \varphi_{,j}
       + {1 \over 3} \chi^i \chi^j \varphi_{,i} \varphi_{,j} \right) \Bigg\}.
   \label{K-bar-eq}
\eea
These equations follow from the Einstein's equation in the ADM (Arnowitt-Deser-Misner) formulation (Arnowitt, Deser \& Misner 1962), presented in equations (A4), (A6)-(A9) of Hwang \& Noh (2013a). These equations without the anisotropic stress are derived in Hwang \& Noh (2013a) and Noh (2014). In order to include the anistropic stress, it is convenient to have the ADM fluid quantities in the following
\bea
   & & E = \widetilde \mu \widehat \gamma^2
         + \widetilde p \left( \widehat \gamma^2 - 1 \right)
         + {1 \over (1 + 2 \varphi )^2} {\widehat v^i \widehat v^j \over c^2} \Pi_{ij}, \quad
         S = 3 \widetilde p + \left( \widehat \mu + \widetilde p \right)
         \left( \widehat \gamma^2 - 1 \right)
         + {1 \over 1 + 2 \varphi} \Pi^i_i,
   \nonumber \\
   & &
         J_i = a \left( \widetilde \mu + \widetilde p \right) \widehat \gamma^2
        {\widehat v_i \over c}
         + {a \over 1 + 2 \varphi} {\widehat v^j \over c} \Pi_{ij}, \quad
         \overline S^i_j
         = \left( \widetilde \mu + \widetilde p \right)
         \left[ {\widehat \gamma^2 \over 1 + 2 \varphi}
         {\widehat v^i \widehat v_j \over c^2}
         - {1 \over 3} \delta^i_j \left( \widehat \gamma^2 - 1 \right) \right]
         + {1 \over 1 + 2 \varphi} \left( \Pi^i_j
         - {1 \over 3} \delta^i_j  \Pi^k_k \right),
\eea
with
\bea
   & & \Pi^i_i
         = {1 \over 1 + 2 \varphi} {\widehat v^i \widehat v^j \over c^2} \Pi_{ij}.
\eea
Equations (\ref{eq1})-(\ref{K-bar-eq}) are valid not only for a single component fluid but also for multi-component system as we regard the fluid quantities as the collective ones. We do not present the (energy and momentum) conservation equations of the collective component because these equations in the presence of the anisotropic stress is rather complicated. This will not be a shortcoming in our case of multiple component fluids and fields system as we will have the (energy and momentum) conservation equations or the equations of motion of individual field component replacing the conservation equations of the collective component. We will consider multiple fluids system without ansiotropic stress of each fluid component, and the minimally coupled scalar field does not have the anisotropic stress. As the multiple nature of fluids and fields generates the anisotropic stress of the collective fluid it is important to keep the anisotropic stress terms in Einstein's equations.

As the dimensions we consider
\bea
   & & [\widetilde g_{ab}] = [\widetilde u_a]
       = [a] = [\gamma_{ij}]
       = [\alpha] = [\varphi] = [\chi^i] = [\chi^{(v)}_i]
       = [\widehat v_i/c] = [\widehat v^{(v)}_i/c]
       = [\widehat \gamma] = 1,
       \nonumber \\
 & &   [x^a] = [ c dt] \equiv [ a d \eta] = L, \quad
       [G \widetilde \varrho] = T^{-2}, \quad
       [\Lambda]
       = [\overline K]
       = L^{-2}, \quad
       [\chi] = [\widehat v/c] = L, \quad
       [\kappa] = T^{-1},
       \nonumber \\
   & & [\widetilde T_{ab}] = [\widetilde \mu]
       = [\widetilde \varrho c^2]
       = [\widetilde p]
       = [\widetilde q_a]
       = [\widetilde \pi_{ab}]
       = [\Pi_{ij}]
       = [\Pi] /L^2
       = [\Pi^{(v)}_i] /L,,
\eea
where $\overline K$ is the normalized background curvature with $R^{(3)} \equiv 6 \overline K$ (Noh 2014).
\end{widetext}

In the above set of equations we have not taken the temporal gauge (hypersurface or slicing) condition yet. As the temporal gauge condition we can impose any one of the following conditions (Bardeen 1988, Hwang 1991, Hwang \& Noh 2013a)
\bea
   & & {\rm comoving \; gauge:}              \hskip 2.13cm     v \equiv 0,
   \nonumber \\
   & & {\rm zero\!-\!shear \; gauge:}        \hskip 1.83cm     \chi \equiv 0,
   \nonumber \\
   & & {\rm uniform\!-\!curvature \; gauge:} \hskip .61cm      \varphi \equiv 0,
   \nonumber \\
   & & {\rm uniform\!-\!expansion \; gauge:} \hskip .57cm      \kappa \equiv 0,
   \nonumber \\
   & & {\rm uniform\!-\!density \; gauge:}   \hskip .97cm      \delta \equiv 0,
   \label{temporal-gauges-NL}
\eea
or combinations of these to all perturbation orders; we can also impose different gauge conditions at different perturbation orders. With the imposition of any of these slicing condition the remaining perturbation variables are free from the remnant (spatial and temporal) gauge mode, and have unique gauge-invariant combinations (Bardeen 1988, Hwang 1991, Noh \& Hwang 2004, Hwang \& Noh 2013a). In the presence of multiple component of fluids and fields we will have additional choice of temporal gauge condition based on the individual component, see equations (\ref{temporal-gauges-I}) and (\ref{temporal-gauges-I-MSFs}).

\section{Multi-component fluids}
                                                    \label{sec:fluids}

Now, we consider the case of multiple components of fluids with vanishing anisotropic stress of each component. We will show that the multiple fluid nature leads to nonvanishing anisotropic stress of the collective fluid. As equations (\ref{eq1})-(\ref{eq5}) are already appended by the anisotropic stress, these are valid with the fluid quantities considered as the collective ones. In the multiple fluids we additionally need to (i) express the collective fluid quantities in terms of the individual ones, and (ii) provide the (energy and momentum) conservation equations for the individual component.

\subsection{Fluid quantities}

In the presence of $N$ fluids we have
\bea
   & & \widetilde T_{ab} = \sum_J \widetilde T_{Jab},
   \label{Tab-Tab-I}
\eea
with the fluid quantities of individual component introduced as
\bea
   & & \widetilde T_{Iab}
       \equiv \widetilde \mu_{I} \widetilde u_{Ia} \widetilde u_{Ib}
       + \widetilde p_{I} \widetilde h_{Iab}
       + \widetilde q_{Ia} \widetilde u_{Ib}
       + \widetilde q_{Ib} \widetilde u_{Ia}
       + \widetilde \pi_{Iab},
   \nonumber \\
\eea
where $\widetilde u_{Ia}$ is the normalized four vectors with $\widetilde u_I^c \widetilde u_{Ic} \equiv -1$, and $\widetilde h_{Iab} \equiv \widetilde g_{ab}       + \widetilde u_{Ia} \widetilde u_{Ib}$ is the projection tensor of each component.
Indices $I, J, \dots = 1, 2 \dots N$ indicate the fluid component. The fluid quantities of individual component can be read as
\bea
   & & \widetilde \mu_I
       = \widetilde T_{Iab} \widetilde u_I^a \widetilde u_I^b, \quad
       \widetilde p_I = {1 \over 3} \widetilde T_{Iab} \widetilde h_I^{ab},
       \nonumber \\
 & &
       \widetilde q_{Ia} = - \widetilde T_{Icd} \widetilde u_I^c
       \widetilde h_{Ia}^{\;\;d}, \quad
       \widetilde \pi_{Iab} = \widetilde T_{Icd} \widetilde h_{Ia}^{\;\;c} \widetilde h_{Ib}^{\;\;d} - \widetilde p_I \widetilde h_{Iab}.
   \nonumber \\
   \label{Tab-decomposition-I}
\eea
Without losing any generality we can take the energy-frame condition $\widetilde q_{Ia} \equiv 0$ for each component.

From equations (\ref{Tab-decomposition}) and (\ref{Tab-decomposition-I}) we have
\bea
   & & \widetilde \mu
       = \sum_J \left\{ \widetilde \mu_{J}
       + \left( \widetilde \mu_{J} + \widetilde p_{J} \right)
       \left[ \left( \widetilde u^c_{J} \widetilde u_c \right)^2 - 1 \right]
       + \widetilde \pi_{Jab} \widetilde u^a \widetilde u^b
       \right\},
   \nonumber \\
   & & \widetilde p
       = \sum_J \Big\{ \widetilde p_{J}
       + {1 \over 3}
       \left( \widetilde \mu_{J} + \widetilde p_{J} \right)
       \left[ \left( \widetilde u^c_{J} \widetilde u_c \right)^2 - 1 \right]
   \nonumber \\
   & & \qquad
       + {1 \over 3} \widetilde \pi_{Jab} \widetilde u^a \widetilde u^b
       \Big\},
   \nonumber \\
   & & \widetilde u_a
       = - {1 \over \widetilde \mu + \sum_K \widetilde p_{K}}
       \sum_J \left[ \left( \widetilde \mu_{J} + \widetilde p_{J} \right)
       \widetilde u_{Ja}
       \widetilde u_{Jb}
       + \widetilde \pi_{Jab} \right] \widetilde u^b,
   \nonumber \\
   & & \widetilde \pi_{ab}
       = \left( \widetilde h^c_a \widetilde h^d_b
       - {1 \over 3} \widetilde h^{cd} \widetilde h_{ab} \right)
       \sum_J \Big[ \left( \widetilde \mu_{J} + \widetilde p_{J} \right)
       \widetilde u_{Jc} \widetilde u_{Jd}
   \nonumber \\
   & & \qquad
       + \widetilde \pi_{Jcd} \Big],
   \label{fluid-Multi-1}
\eea
thus
\bea
   & &
       \widetilde \mu - 3 \widetilde p
       = \sum_J \left( \widetilde \mu_J
       - 3 \widetilde p_J \right).
\eea In the following we ignore $\widetilde \pi_{Icd}$.

We introduce the fluid four-vector of individual component as \bea
   & & \widetilde u_{Ii} \equiv {a \widehat \gamma_I \widehat v_{Ii} \over c},
\eea
thus we have
\bea
   & & \widetilde u_{I0} = - \widehat \gamma_I \left( a {\cal N}
        +{\widehat v_{Ik} \chi^{k} \over c (1 + 2 \varphi)} \right),
   \nonumber \\
   & &
        \widetilde u_I^i = {\widehat \gamma_I \over a (1 + 2 \varphi)}
       \left( {\widehat v_I^i \over c}
       + {\chi^i \over a {\cal N}} \right), \quad
       \widetilde u_I^0
       = {1 \over a {\cal N}} \widehat \gamma_I,
   \nonumber \\
   & &
       \widehat \gamma_I
       \equiv {1 \over \sqrt{ 1 - {\widehat v_I^k \widehat v_{Ik}
       \over c^2 (1 + 2 \varphi)}}},
   \label{four-vector-I}
\eea
where spatial index of $\widehat v_{Ii}$ is raised and lowered by $\gamma_{ij}$ as the metric. For the collective component, we can simply remove the subindex indicating the component.

\begin{widetext}
The fluid quantities in equation (\ref{fluid-Multi-1}) give
\bea
   & & \widetilde \mu
       = \sum_J \left\{ \widetilde \mu_J
       + \left( \widetilde \mu_J + \widetilde p_J \right)
       \left[ \widehat \gamma^2 \widehat \gamma_J^2
       \left( 1- {1 \over 1 + 2 \varphi} {\widehat v^k \widehat v_{Jk} \over c^2} \right)^2 - 1 \right] \right\},
   \nonumber \\
   & & \widetilde p
       = \sum_J \left\{ \widetilde p_J
       + {1 \over 3} \left( \widetilde \mu_J + \widetilde p_J \right)
       \left[ \widehat \gamma^2 \widehat \gamma_J^2
       \left( 1- {1 \over 1 + 2 \varphi} {\widehat v^k \widehat v_{Jk} \over c^2} \right)^2 - 1 \right] \right\},
   \nonumber \\
   & & \widehat v_i
       = {1 \over \widehat \gamma^2}
       { \sum_J \left( \widetilde \mu_{J} + \widetilde p_{J} \right)
       \widehat \gamma_J^2
       \left( 1 - {1 \over 1 + 2 \varphi}
       {\widehat v_j \widehat v_J^j \over c^2} \right) \widehat v_{Ji}
       \over
       \sum_K \left( \widetilde \mu_K + \widetilde p_K \right)
       \widehat \gamma_K^2
       \left( 1 - {1 \over 1 + 2 \varphi}
       {\widehat v_k \widehat v_K^k \over c^2} \right)^2 },
   \nonumber \\
   & & \Pi_{ij}
       = \sum_J \left( \widetilde \mu_J + \widetilde p_J \right)
       \Bigg\{
       \widehat \gamma_J^2
       \left[ \widehat \gamma^2
       \left( 1- {1 \over 1 + 2 \varphi} {\widehat v^k \widehat v_{Jk} \over c^2} \right) {\widehat v_i \over c}
       - {\widehat v_{Ji} \over c} \right]
        \left[ \widehat \gamma^2
       \left( 1- {1 \over 1 + 2 \varphi} {\widehat v^\ell \widehat v_{J\ell} \over c^2} \right) {\widehat v_j \over c}
       - {\widehat v_{Jj} \over c} \right]
   \nonumber \\
   & & \qquad
       - {1 \over 3} \left[ \left( 1 + 2 \varphi \right) \gamma_{ij}
       + \widehat \gamma^2 {\widehat v_i \widehat v_j \over c^2} \right]
       \left[ \widehat \gamma^2 \widehat \gamma_J^2
       \left( 1- {1 \over 1 + 2 \varphi} {\widehat v^k \widehat v_{Jk} \over c^2} \right)^2 - 1 \right]
       \Bigg\}.
   \label{fluid-Multi-2}
\eea
Equation (\ref{fluid-Multi-2}) provides the relations between the collective and individual fluid quantities. Using these relations equations (\ref{eq1})-(\ref{eq5}) remain valid even in the multiple component fluid case. As the right-hand-side of $\widehat v_i$ in equation (\ref{fluid-Multi-2}) still has $\widehat v_k$, in order to express $\widehat v_i$ in terms of individual fluid component only, we need to solve this relation perturbatively by iteration. The other collective fluid quantities also contain $\widehat v_i$ which should be replaced by thus obtained individual ones.

For example, to the second order, we have \bea
   & & \widehat v_i
       = {\sum_J \left( \widetilde \mu_J + \widetilde p_J \right) \widehat v_{Ji}
       \over \sum_K \left( \widetilde \mu_K + \widetilde p_K \right)},
   \label{v-v-2nd}
\eea
and to the third order, we have
\bea
   & & \widetilde v_i
       = {\sum_J \left( \widetilde \mu_J + \widetilde p_J \right)
       \left[ 1 + {1 \over c^2} \left(
       \widehat v_k \widehat v^k
       - \widehat v_k \widehat v_J^k
       + \widehat v_{Jk} \widehat v_J^k \right) \right]
       \widehat v_{Ji}
       \over
       \sum_K \left( \widetilde \mu_K + \widetilde p_K \right)
       \left[ 1 + {1 \over c^2} \widehat v_{K\ell}
       \left( \widehat v_K^\ell - 2 \widehat v^\ell \right) \right] },
\eea
where $\widehat v_j$'s in the right-hand-side can be replaced by the one in equation (\ref{v-v-2nd}). This can be continued to higher order perturbations. The rest of fluid quantities to the third order are
\bea
   & & \widetilde \mu
       = \sum_J \left[ \widetilde \mu_J
       + {\widetilde \mu_J + \widetilde p_J \over 1 + 2 \varphi}
       {1 \over c^2} \left( \widehat v^k - \widehat v_J^k \right)
       \left( \widehat v_k - \widehat v_{Jk} \right) \right], \quad
       \widetilde p
       = \sum_J \left[ \widetilde p_J
       + {1 \over 3} {\widetilde \mu_J + \widetilde p_J \over 1 + 2 \varphi}
       {1 \over c^2} \left( \widehat v^k - \widehat v_J^k \right)
       \left( \widehat v_k - \widehat v_{Jk} \right) \right],
   \nonumber \\
   & & \Pi_{ij}
       = \sum_J \left( \widetilde \mu_J + \widetilde p_J \right)
       {1 \over c^2} \left[ \left( \widehat v_i - \widehat v_{Ji} \right)
       \left( \widehat v_j - \widehat v_{Jj} \right)
       - {1 \over 3} \gamma_{ij}
       \left( \widehat v^k - \widehat v_J^k \right)
       \left( \widehat v_k - \widehat v_{Jk} \right) \right],
\eea
where $\widehat v_j$ in the right-hand-sides can be replaced by the one in equation (\ref{v-v-2nd}).

\subsection{Conservation equations of individual component}

Now, we provide the energy and momentum conservation equations followed by the individual fluid. The energy and momentum conservation equations follow from $\widetilde T^b_{a;b} = 0$, thus
\bea
   & & \widetilde T^{\;\; b}_{Ia;b} \equiv \widetilde I_{Ia}, \quad
       \sum_J \widetilde I_{Ja} = 0,
\eea
where $\widetilde I_{Ia}$ indicates the interaction terms among fluids.

The ADM and the covariant (energy and momentum) conservation equations for individual component are presented in equations (E10)-(E13) of Hwang \& Noh (2013a). Using the ADM and the covariant quantities presented in the Appendices B and C of Hwang \& Noh (2013a) we can derive these equations in the fully nonlinear forms as the following.

\noindent
ADM energy conservation:
\bea
   & & {1 \over {\cal N}} \left[ \widetilde \mu_I
       + \left( \widetilde \mu_I + \widetilde p_I \right)
       \left( \widehat \gamma_I^2 - 1 \right) \right]^{\displaystyle\cdot}
       + {c \over a^2 {\cal N}} {\chi^i \over 1 + 2 \varphi}
       \left[ \widetilde \mu_I
       + \left( \widetilde \mu_I +\widetilde p_I \right)
       \left( \widehat \gamma_I^2 - 1 \right) \right]_{|i}
   \nonumber \\
   & & \qquad
       + \left( \widetilde \mu_I + \widetilde p_I \right) (3 H - \kappa)
       {1 \over 3} \left( 4 \widehat \gamma_I^2 - 1 \right)
       + \left( {\widetilde \mu_I + \widetilde p_I
       \over a ( 1 + 2 \varphi )} \widehat \gamma_I^2 \widehat v_I^i \right)_{|i}
       +\left( {3 \varphi_{,i} \over 1 + 2 \varphi}
       + 2 {{\cal N}_{,i} \over {\cal N}} \right)
       {\widetilde \mu_I + \widetilde p_I \over a ( 1 + 2 \varphi )}
       \widehat \gamma_I^2 \widehat v_I^i
   \nonumber \\
   & & \qquad
       + {\widehat \gamma_I^2 (\widetilde \mu_I + \widetilde p_I)
       \over c a^2 {\cal N} ( 1 + 2 \varphi)^2}
       \left[ \chi^{i|j} \widehat v_{Ii} \widehat v_{Ij}
       - {1 \over 3} \chi^j_{\;\; |j} \widehat v_I^i \widehat v_{Ii}
       - {2 \over 1 + 2 \varphi}
       \left( \widehat v_I^i \widehat v_I^j \chi_i \varphi_{,j}
       - {1 \over 3} \widehat v_I^i \widehat v_{Ii} \chi^j \varphi_{,j} \right)
       \right]
   \nonumber \\
   & & \qquad
       =
       - {c \over a {\cal N}}
       \left( \widetilde I_{I0}
       + {\chi^i \over a (1 + 2 \varphi)} I_{Ii} \right).
   \label{eq6-ADM}
\eea

\noindent
ADM momentum conservation:
\bea
   & & \left( {1 \over {\cal N}}
       {\partial \over \partial t}
       + 3 H -\kappa \right)
       \left[ a (\widetilde \mu_I + \widetilde p_I ) \widehat \gamma_I^2
       \widehat v_{Ii} \right]
       + {c \over a^2 {\cal N}} {\chi^j \over 1 + 2 \varphi}
       \left[ a (\widetilde \mu_I + \widetilde p_I)
       \widehat \gamma_I^2 \widehat v_{Ii} \right]_{|j}
       + c^2 \widetilde p_{I,i}
       + c^2 \left( \widetilde \mu_I + \widetilde p_I \right)
       {{\cal N}_{,i} \over {\cal N}}
   \nonumber \\
   & & \qquad
       + \left( {\widetilde \mu_I + \widetilde p_I \over 1 + 2 \varphi}
       \widehat \gamma_I^2 \widehat v_I^j \widehat v_{Ii} \right)_{|j}
       + {c \over a {\cal N}} \left( {\chi^j \over 1 + 2 \varphi} \right)_{|i}
       \left( \widetilde \mu_I + \widetilde p_I \right)
       \widehat \gamma_I^2 \widehat v_{Ij}
   \nonumber \\
   & & \qquad
       + {\widetilde \mu_I + \widetilde p_I \over 1 + 2 \varphi}
       \widehat \gamma_I^2 \widehat v_I^j
       \left[ {1 \over 1 + 2 \varphi} (3 \widehat v_{Ii} \varphi_{,j}
       - \widehat v_{Ij} \varphi_{,i} )
       + {1 \over {\cal N}} ( \widehat v_{Ii} {\cal N}_{,j}
       + \widehat v_{Ij} {\cal N}_{,i} ) \right]
       = c^2 I_{Ii}.
   \label{eq7-ADM}
\eea

\noindent
Covariant energy conservation:
\bea
   & & \left[ {\partial \over \partial t}
       + {1 \over a (1 + 2 \varphi)}
       \left( {\cal N} \widehat v_I^i + {c \over a} \chi^i \right)
       \nabla_i \right] \widetilde \mu_I
       + \left( \widetilde \mu_I + \widetilde p_I \right)
       \Big\{ \left( 3 H - \kappa \right) {\cal N}
       + { ( {\cal N} \widehat v_I^i )_{|i} \over a ( 1 + 2 \varphi)}
       +{ {\cal N} \widehat v_I^i \varphi_{,i} \over a ( 1 + 2 \varphi )^2}
   \nonumber \\
   & & \qquad
       + {1 \over \widehat \gamma_I}
       \left[ {\partial \over \partial t}
       + {1 \over a ( 1 + 2 \varphi )}
       \left( {\cal N} \widehat v_I^i + {c \over a} \chi^i \right)
       \nabla_i \right] \widehat \gamma_I
       \Big\}
       = - {c \over a} \widetilde I_{I0}
       - {{\cal N} \over a (1 + 2 \varphi)}
       \left( \widehat v_I^i
       + {c \over a {\cal N}} \chi^i \right)
       I_{Ii}.
   \label{eq6-cov}
\eea

\noindent
Covariant momentum conservation:
\bea
   & & {\partial \over \partial t}
       \left( a \widehat \gamma_I \widehat v_{Ii} \right)
       + {1 \over a (1 + 2 \varphi)}
       \left( {\cal N} \widehat v_I^k + {c \over a} \chi^k \right)
       \nabla_k \left( a \widehat \gamma_I \widehat v_{Ii} \right)
       + c^2 \widehat \gamma_I {\cal N}_{,i}
       + {1 - \widehat \gamma_I^2 \over \widehat \gamma_I}
       {c^2 {\cal N} \varphi_{,i} \over 1 + 2 \varphi}
       + {c \over a} \widehat \gamma_I \widehat v_I^k
       \nabla_i \left( {\chi_k \over 1 + 2 \varphi} \right)
   \nonumber \\
   & & \qquad
       + {1 \over \widetilde \mu_I + \widetilde p_I}
       \left\{ c^2 {{\cal N} \over \widehat \gamma_I} \widetilde p_{I,i}
       +a \widehat \gamma_I \widehat v_{Ii}
       \left[ {\partial \over \partial t}
       + {1 \over a (1 + 2 \varphi)}
       \left( {\cal N} \widehat v_I^k
       + {c \over a} \chi^k \right) \nabla_k \right]
       \widetilde p_I \right\}
   \nonumber \\
   & & \qquad
       = {c^2 \over \widetilde \mu_I + \widetilde p_I}
       \left[ {{\cal N} \over \widehat \gamma_I} I_{Ii}
       + \widehat \gamma_I {\widehat v_{Ii} \over c} \widetilde I_{I0}
       + {{\cal N} \widehat \gamma_I \over 1 + 2 \varphi}
       {\widehat v_{Ii} \over c}
       \left( {\widehat v_I^j \over c}
       + {\chi^j \over a {\cal N}} \right) I_{Ij} \right].
   \label{eq7-cov}
\eea
We have introduced $I_{Ii} \equiv \widetilde
I_{Ii}$ where the spatial index of $I_{Ii}$ is raised and
lowered by $\gamma_{ij}$; $I_{Ii}$ is the perturbed order quantity
whereas $\widetilde I_{0}$ includes the background order
quantity as $\widetilde I_{I0} = I_{I0} + \delta I_{I0}$.
Either set of conservation equations together with Eqs. (\ref{eq1})-(\ref{eq5}) complete the equations we need in the multiple fluid system; equation
(\ref{fluid-Multi-2}) provides the collective fluid quantities in terms of the individual ones.
\end{widetext}

The vector variables $v_{Ii}$ and
$I_{Ii}$ can be decomposed into the scalar- and vector-type
perturbations as \bea
   & & v_{Ii} = - v_{I,i} + v^{(v)}_{Ii}, \quad
       I_{Ii} = I_{I,i} + I^{(v)}_{Ii},
\eea
with $v^{(v)i}_{I\;\;\;\;\;|i} \equiv 0 \equiv I^{(v)i}_{I\;\;\;\;\;|i}$.

As in the single component case, to the linear order the scalar-type perturbation $\delta_{I}$, $\delta p_{I}$, $v_{I}$, $\delta I_{I0}$ and $\delta I_{I}$ depend on the temporal gauge transformation whereas the vector-type perturbations are gauge invariant (Bardeen 1988). In addition to the fundamental gauge conditions in equation (\ref{temporal-gauges-NL}), in the multi-component case, for a chosen $I$-component, we have the following gauge conditions available
\bea
   & & I{\rm - component\!-\!comoving \; gauge:} \hskip 1.5cm v_{I} \equiv 0,
   \nonumber \\
   & & {\rm uniform\!}-\!I\!-{\rm \!component\!-\!density \; gauge:} \hskip .3cm \delta_{I} \equiv 0,
   \nonumber \\
   \label{temporal-gauges-I}
\eea
to the fully nonlinear order.

\section{Newtonian Limit}
                                                \label{sec:Newtonian}

In Hwang \& Noh (2013b) we have shown the Newtonian theory as the infinite-speed-of-light limit. The Newtonian limit is available in both the zero-shear gauge and the uniform-expansion gauge. In this section we will present the Newtonian limit in multiple component fluids in the same gauge conditions. As in Hwang \& Noh (2013b) we will {\it assume} a flat background, and {\it ignore} the anisotropic stress of the individual fluid. Even in this Newtonian limit we will show that anisotropic stress of collective fluid is generated from the multiple component nature of the fluids system.

In both gauge conditions, the infinite-speed-of-light limit ($c \rightarrow \infty$) implies
\bea
   & & \alpha \ll 1, \quad
       \varphi \ll 1, \quad
       {\widehat v_I^i \widehat v_{Ii} \over c^2} \ll 1,
   \nonumber \\
   & &
       {a^2 H^2 \over k^2 c^2} \ll 1, \quad
       {\widetilde p_I \over \widetilde \varrho_I c^2} \ll 1,
   \label{Newtonian-limit}
\eea
where $k$ is the wavenumber with $\Delta \rightarrow - k^2$, $H \equiv \dot a/a$, and $\widetilde \mu_I \equiv \widetilde \varrho_I c^2$. The first two conditions are the weak-gravity conditions, and the remaining ones are the slow-motion, the small-scale (smaller than the horizon), and negligible pressure conditions, respectively. As we consider the infinite-speed-of-light limit, we will ignore the above quantities but we will keep $c^2 \widetilde p_{I,i}$.

Equation (\ref{eq3}) gives
\bea
   & & {2 \over 3} \kappa_{,i}
       + c {\Delta \over a^2 {\cal N}} \left( {2 \over 3} \chi_{,i}
       + {1 \over 2} \chi^{(v)}_i \right)
   \nonumber \\
   & & \qquad
       - {c {\cal N}_{,j} \over a^2 {\cal N}^2}
       \left[ \chi^{,j}_{\;\;\; i} - {1 \over 3} \delta^j_i \Delta \chi
       + {1 \over 2} \left( \chi^{(v)j}_{\;\;\;\;\;\; ,i}
       + \chi^{(v),j}_i \right) \right]
   \nonumber \\
   & & \qquad
       = - {8 \pi G \over c^2} a \widetilde \varrho
       \left( - \widehat v_{,i} + \widehat v^{(v)}_i \right),
   \label{eq3-N}
\eea with
\bea
   & & {\cal N} = \sqrt{ 1 + {1 \over a^2} ( \chi^{,k} + \chi^{(v)k} )
        ( \chi_{,k} + \chi^{(v)}_k ) },
\eea
where we decompose into scalar- and vector-type perturbations using equation (\ref{s-v}). Here we have properly considered the anisotropic stress of the collective component which arises from the multi-component nature of the fluids, see equation (\ref{fluids-N}). This equation can be analyzed perturbatively in the two gauge conditions.

In the zero-shear gauge ($\chi \equiv 0$), in a perturbative manner we can show
\bea
   & & {1 \over a} \chi^{(v)}_i
       \sim {\cal O} \left( {a^2 H^2 \over k^2 c^2} {\widehat v_i \over c} \right),
\eea
to all perturbation order, thus $\chi_i^{(v)} / a$ is negligible in our limit. Thus, to fully nonlinear order equation (\ref{eq3-N}) can be written as
\bea
   & & {2 \over 3} \kappa_{,i}
       + {c \Delta \over 2 a^2} \chi^{(v)}_i
       = - {8 \pi G \over c^2} a \widetilde \varrho
       \widehat v_i,
\eea
and, we have
\bea
   & & \kappa
       = - {12 \pi G a \over c^2} \Delta^{-1} \nabla_i
       \left( \widetilde \varrho \widehat v^i \right),
   \nonumber \\
   & &
       \chi^{(v)}_i
       = - {16 \pi G \over c^3} a^3 \Delta^{-1}
       \left[ \widetilde \varrho \widehat v_i
       - \Delta^{-1} \nabla_i \nabla_j \left( \widetilde \varrho \widehat v^j
       \right) \right].
   \label{kappa-ZSG}
\eea

In the uniform-expansion gauge ($\kappa \equiv 0$), in a perturbative manner we can show
\bea
   & & {1 \over a} \chi_{,i}
       \sim {1 \over a} \chi^{(v)}_i
       \sim {\cal O} \left( {a^2 H^2 \over k^2 c^2} {\widehat v_i \over c} \right),
\eea
to all perturbation orders, thus $\chi_i / a$ is negligible in our limit. Thus, to fully nonlinear order equation (\ref{eq3-N}) can be written as
\bea
   & & \chi_{,i} + {3 \over 4} \chi^{(v)}_i
       = - {12 \pi G \over c^3} a^3 \Delta^{-1}
       \left( \widetilde \varrho \widehat v_i \right),
\eea
and, we have
\bea
   & & \chi
       = - {12 \pi G \over c^3} a^3 \Delta^{-2} \nabla_i
       \left( \widetilde \varrho \widehat v^i \right),
   \nonumber \\
   & &
       \chi^{(v)}_i
       = - {16 \pi G \over c^3} a^3 \Delta^{-1} \left[
       \widetilde \varrho \widehat v_i
       - \Delta^{-1} \nabla_i \nabla_j
       \left( \widetilde \varrho \widehat v^j \right) \right].
   \label{chi-UEG}
\eea

Thus we have shown that for $c \rightarrow \infty$ we have
\bea
   & & \chi^{(v)}_i = 0,
\eea
in both gauge conditions. This does not imply that we ignore the vector-type perturbation. We will keep $\widehat v^{(v)}_i$ in $\widehat v_i$. As $\chi_i/a$ is negligible we have ${\cal N} = 1 + \alpha$ in our limit, and we will keep $c^2 {\cal N}_{,i} = c^2 \alpha_{,i}$ in the momentum conservation equation.

In the zero-shear gauge ($\chi \equiv 0$) we have $\chi_i = 0$. Using $\kappa$ in equation (\ref{kappa-ZSG}), equations (\ref{eq1}), (\ref{eq2}), (\ref{eq4}) and (\ref{eq5}), respectively, give
\bea
   & & \dot \varphi - {\dot a \over a} \alpha
       = {4 \pi G a \over c^2} \Delta^{-1} \nabla_i
       \left( \widetilde \varrho \widehat v^i \right),
   \nonumber \\
   & &
       4 \pi G \delta \varrho + c^2 {\Delta \over a^2} \varphi = 0,
   \nonumber \\
   & &
       4 \pi G \delta \varrho - c^2 {\Delta \over a^2} \alpha = 0,
   \nonumber \\
   & &
       \varphi = - \alpha.
   \label{ZSG-relations}
\eea
The energy conservation and the momentum conservation equations in equations (\ref{eq6-ADM})-(\ref{eq7-cov}) give
\bea
   & & \dot {\widetilde \varrho}_I
       + 3 {\dot a \over a} \widetilde \varrho_I
       + {1 \over a} \nabla_i \left( \widetilde \varrho_I \widehat v_I^i \right)
       = 0,
   \label{E-conserv-ZSG} \\
   & & \dot {\widehat v}_{Ii}
       + {\dot a \over a} \widehat v_{Ii}
       + {1 \over a} \widehat v_I^k \nabla_k \widehat v_{Ii}
       = - {c^2 \over a} \alpha_{,i}
       - {\widetilde p_{I,i} \over a \widetilde \varrho_I}.
   \label{mom-conserv-ZSG}
\eea

In the uniform-expansion gauge ($\kappa \equiv 0$) we have $\chi_i = \chi_{,i}$. Using $\chi$ in equation (\ref{chi-UEG}), we can show that the same equations used in the zero-shear gauge above lead to exactly the same equations in equations (\ref{ZSG-relations})-(\ref{mom-conserv-ZSG}). Thus the following analysis applies for both gauges.

Fluid quantities in equation (\ref{ZSG-relations}) are collective ones. The relations between individual fluid and collective one are presented in equation (\ref{fluid-Multi-2}). In the $c \rightarrow \infty$ limit, we have
\bea
   & & \widetilde \varrho = \sum_K \widetilde \varrho_K, \quad
       \widetilde p = \sum_K \widetilde p_K, \quad
       \widetilde \varrho \widehat v_i
       = \sum_K \widetilde \varrho_K \widehat v_{Ki},
   \nonumber \\
   & &
       \Pi_{ij}
       = - \widetilde \varrho \widehat v_i \widehat v_j
       + \sum_K \widetilde \varrho_K \widehat v_{Ki} \widehat v_{Kj}.
   \label{fluids-N}
\eea Notice that we have non-vanishing anisotropic stress of the collective fluid generated from the multiple nature of fluid system.

Now, we identify the Newtonian densities, pressures and velocities of individual component, and the gravitational potential as
\bea
   & & \widetilde \varrho, \quad
       \widetilde \varrho_I, \quad
       \widetilde p, \quad
       \widetilde p_I, \quad
       {\bf v} = v^i \equiv \widehat v^i, \quad
       {\bf v}_I = v_I^i \equiv \widehat v_I^i,
   \nonumber \\
   & &
       {1 \over c^2} U \equiv - \alpha = \varphi.
   \label{N-identification}
\eea Using these variables, equations (\ref{ZSG-relations})-(\ref{mom-conserv-ZSG}) give
\bea
   & & \dot {\widetilde \varrho}_I
       + 3 {\dot a \over a} \widetilde \varrho_I
       + {1 \over a} \nabla \cdot \left( \widetilde \varrho_I {\bf v}_I \right)
       = 0,
   \label{mass-conserv-N-I} \\
   & & \dot {\bf v}_I
       + {\dot a \over a} {\bf v}_I
       + {1 \over a} {\bf v}_I \cdot \nabla {\bf v}_I
       =  {1 \over a} \nabla U
       - {1 \over a \widetilde \varrho_I} \nabla \widetilde p_I.
   \label{mom-conserv-N-I} \\
   & & 4 \pi G \delta \varrho = - {\Delta \over a^2} U,
   \label{Poisson-eq-N}
\eea which are the same as the mass conservation, momentum conservation, and the Poisson's equations, respectively, in Newtonian context.

For the collective fluid we can show that
\bea
   & & \dot {\widetilde \varrho}
       + 3 {\dot a \over a} \widetilde \varrho
       + {1 \over a} \nabla \cdot \left( \widetilde \varrho {\bf v} \right)
       = 0,
   \label{mass-conserv-N} \\
   & & \dot {\bf v}
       + {\dot a \over a} {\bf v}
       + {1 \over a} {\bf v} \cdot \nabla {\bf v}
       =  {1 \over a} \nabla U
       - {1 \over a \widetilde \varrho} \left(
       \nabla \widetilde p
       + \nabla^j \Pi_{ij} \right),
   \nonumber \\
   \label{mom-conserv-N}
\eea
with
\bea
   & & \widetilde \varrho = \sum_K \widetilde \varrho_K, \quad
       \widetilde p = \sum_K \widetilde p_K, \quad
       \widetilde \varrho {\bf v}
       = \sum_K \widetilde \varrho_K {\bf v}_K,
   \nonumber \\
   & &
       \Pi_{ij}
       = - \widetilde \varrho v_i v_j
       + \sum_K \widetilde \varrho_K v_{Ki} v_{Kj}.
\eea

We still have one more relation to check. Using the identification in equation (\ref{N-identification}) the first equation in equation (\ref{ZSG-relations}) gives
\bea
   & & \dot U + {\dot a \over a} U
       = 4 \pi G a \Delta^{-1} \nabla \cdot
       \left( \widetilde \varrho {\bf v} \right),
   \label{U-eq}
\eea
which follows from equations (\ref{Poisson-eq-N}) and (\ref{mass-conserv-N}). In the presence of relativistic pressure, see the paragraph including equation (\ref{discrepancy}). In our infinite-speed-of-light limit we have checked that the complete set of Einstein's equations is consistent.

This completes the proof of the Newtonian limit of Einstein's gravity in the cosmological context. By setting $a = 1$ with $\varrho = 0$, thus $\delta \varrho = \widetilde \varrho$, we properly recover the well known Newtonian limit in the Minkowski background.

\section{Newtonian limit with relativistic pressure}

In Hwang \& Noh (2013c) we have presented Newtonian equations with general relativistic pressure, thus $\widetilde p \sim \widetilde \varrho c^2$, for a single component fluid in the zero-shear gauge. It is not necessary that such a limit should exist at all, but we have shown that such a limit does exist with correct special relativistic hydrodynamics limit for vanishing gravity. Here, we extend the case to the multi-component fluid system.

We take the zero-shear gauge, thus set $\chi \equiv 0$. We impose the infinite-speed-of-light conditions in equation (\ref{Newtonian-limit}) except for the last one concerning the pressure. Thus, we consider a situation with $\widetilde p_I \sim \widetilde \varrho_I c^2$. The analysis proceeds similarly as in the Newtonian limit studied in Sec.\ \ref{sec:Newtonian}.

Equation (\ref{eq3}) gives
\bea
   & & {2 \over 3} \kappa_{,i}
       + c {\Delta \over a^2 {\cal N}}
       {1 \over 2} \chi^{(v)}_i
       - {c {\cal N}_{,j} \over a^2 {\cal N}^2}
       {1 \over 2} \left( \chi^{(v)j}_{\;\;\;\;\;\; ,i}
       + \chi^{(v),j}_i \right)
   \nonumber \\
   & & \qquad
       = - {8 \pi G \over c^2} a \left( \widetilde \varrho
       + {\widetilde p \over c^2} \right)
       \widehat v_i,
   \label{eq3-N-pressure}
\eea with
\bea
   & & {\cal N} = \sqrt{ 1 + {1 \over a^2} \chi^{(v)k} \chi^{(v)}_k }.
\eea
In a perturbative manner we can show
\bea
   & & {1 \over a} \chi^{(v)}_i
       \sim {\cal O} \left( {a^2 H^2 \over k^2 c^2} {\widehat v_i \over c} \right),
\eea
to all perturbation orders, thus $\chi_i^{(v)} / a$ is negligible in our limit. Thus, to fully nonlinear order equation (\ref{eq3-N-pressure}) becomes
\bea
   & & {2 \over 3} \kappa_{,i}
       + {c \Delta \over 2 a^2} \chi^{(v)}_i
       = - {8 \pi G \over c^2} a \left( \widetilde \varrho
       + {\widetilde p \over c^2} \right)
       \widehat v_i,
\eea
and, we have
\bea
   & & \kappa
       = - {12 \pi G a \over c^2} \Delta^{-1} \nabla_i
       \left[ \left( \widetilde \varrho
       + {\widetilde p \over c^2} \right) \widehat v^i \right],
   \nonumber \\
   & &
       \chi^{(v)}_i
       = - {16 \pi G \over c^3} a^3 \Delta^{-1}
       \left\{ \left( \widetilde \varrho
       + {\widetilde p \over c^2} \right) \widehat v_i \right.
   \nonumber \\
   & & \qquad
       \left.
       - \Delta^{-1} \nabla_i \nabla_j \left[ \left( \widetilde \varrho
       + {\widetilde p \over c^2} \right) \widehat v^j
       \right] \right\}.
   \label{kappa-ZSG-pressure}
\eea
Thus, we have
\bea
   & & \chi^{(v)}_i = 0,
\eea
but we keep $\widehat v^{(v)}_i$ in $\widehat v_i$. As $\chi_i/a$ is negligible we have ${\cal N} = 1 + \alpha$ with $\alpha \ll 1$; thus we have ${\cal N} = 1$, but we will keep $c^2 {\cal N}_{,i} = c^2 \alpha_{,i}$ in the momentum conservation equation.

Using $\kappa$ in equation (\ref{kappa-ZSG-pressure}), equations (\ref{eq1}), (\ref{eq2}), (\ref{eq4}) and (\ref{eq5}), respectively, give
\bea
   & & \dot \varphi - {\dot a \over a} \alpha
       = {4 \pi G a \over c^2} \Delta^{-1} \nabla_i
       \left[ \left( \widetilde \varrho
       + {\widetilde p \over c^2} \right) \widehat v^i \right],
   \label{pressure-relation1} \\
   & &
       4 \pi G \delta \varrho + c^2 {\Delta \over a^2} \varphi
       = - {\dot a \over a} \kappa,
   \label{pressure-relation2} \\
   & &
       4 \pi G \left( \delta\varrho
       + 3 {\delta p \over c^2} \right)
       - c^2 {\Delta \over a^2} \alpha
       = \dot \kappa + 2 {\dot a \over a} \kappa,
   \label{pressure-relation3} \\
   & &
       \varphi = - \alpha.
   \label{pressure-relation4}
\eea
Although $\kappa$ is higher order in $c^{-2}$ it is important to keep this term because $\dot \kappa$ is no longer higher order in the presence of relativistic pressure, see equation (\ref{dot-kappa-pressure}). Fluid quantities in equations (\ref{pressure-relation1})-(\ref{pressure-relation3}) are collective ones. From equation (\ref{fluid-Multi-2}) we have
\bea
   & & \widetilde \varrho = \sum_K \widetilde \varrho_K, \quad
       \widetilde p = \sum_K \widetilde p_K,
   \nonumber \\
   & &
       \left( \widetilde \varrho
       + {\widetilde p \over c^2} \right) \widehat v_i
       = \sum_K \left( \widetilde \varrho_K
       + {\widetilde p_K \over c^2} \right) \widehat v_{Ki},
   \nonumber \\
   & &
       \Pi_{ij}
       = - \left( \widetilde \varrho
       + {\widetilde p \over c^2} \right) \widehat v_i \widehat v_j
       + \sum_K \left( \widetilde \varrho_K
       + {\widetilde p_K \over c^2} \right) \widehat v_{Ki} \widehat v_{Kj}.
   \nonumber \\
   \label{fluids-N-pressure}
\eea

The energy conservation and the momentum conservation equations in equations (\ref{eq6-ADM})-(\ref{eq7-cov}) give
\bea
   & & \dot {\widetilde \varrho}_I
       + 3 {\dot a \over a} \left( \widetilde \varrho_I
       + {\widetilde p_I \over c^2} \right)
       + {1 \over a} \nabla_i \left( \widetilde \varrho_I \widehat v_I^i \right)
   \nonumber \\
   & & \qquad
       = {1 \over a c^2} \left( \widehat v_I^i \nabla_i \widetilde p_I
       - \widetilde p_I \nabla_i \widehat v_I^i \right),
   \label{E-conserv-I-pressure} \\
   & & \dot {\widehat v}_{Ii}
       + {\dot a \over a} \widehat v_{Ii}
       + {1 \over a} \widehat v_I^k \nabla_k \widehat v_{Ii}
       = - {c^2 \over a} \alpha_{,i}
   \nonumber \\
   & & \qquad
       - {1 \over \widetilde \varrho_I + \widetilde p_I/ c^2}
       \left( {1 \over a} \nabla_i \widetilde p_I
       + \widehat v_{Ii} {\dot {\widetilde p}_I \over c^2} \right).
   \label{mom-conserv-I-pressure}
\eea
We identify the Newtonian densities, pressures and velocities of individual component, and the gravitational potential, which are the same as those in equation (\ref{N-identification}) in the Newtonian limit. Using the Newtonian variables, equations (\ref{E-conserv-I-pressure}) and (\ref{mom-conserv-I-pressure}) become
\bea
   & & \dot {\widetilde \varrho}_I
       + 3 {\dot a \over a} \left( \widetilde \varrho_I
       + {\widetilde p_I \over c^2} \right)
       + {1 \over a} \nabla \cdot \left[ \left( \widetilde \varrho_I
       + {\widetilde p_I \over c^2} \right) {\bf v}_I \right]
   \nonumber \\
   & & \qquad
       = {2 \over a c^2} {\bf v}_I \cdot \nabla \widetilde p_I,
   \label{E-conserv-I-pressure-N} \\
   & & \dot {\bf v}_I
       + {\dot a \over a} {\bf v}_I
       + {1 \over a} {\bf v}_I \cdot \nabla {\bf v}_I
   \nonumber \\
   & & \qquad
       = {1 \over a} \nabla U
       - {1 \over \widetilde \varrho_I + \widetilde p_I/ c^2}
       \left( {1 \over a} \nabla \widetilde p_I
       + {\bf v}_I {\dot {\widetilde p}_I \over c^2} \right).
   \label{mom-conserv-I-pressure-N}
\eea

Now, we derive the Poisson's equation. From the $\kappa$ relation in equation (\ref{kappa-ZSG-pressure}), using equations (\ref{E-conserv-I-pressure-N}) and (\ref{mom-conserv-I-pressure-N}), we can show
\bea
   & & \dot \kappa = {12 \pi G \over c^2} \delta p.
   \label{dot-kappa-pressure}
\eea
Thus, as we mentioned, in our limit it is important to keep $\dot \kappa$ term. Using this, the Poisson's equation follows from equations (\ref{pressure-relation2})-(\ref{pressure-relation4}) as
\bea
   & & 4 \pi G \delta \varrho
       = - {\Delta \over a^2} U.
   \label{Poisson-eq-pressure}
\eea
Equations (\ref{E-conserv-I-pressure-N}), (\ref{mom-conserv-I-pressure-N}) and (\ref{Poisson-eq-pressure}) are the mass and momentum conservation equations for the individual fluid component, and the Poisson's equation in the presence of the general relativistic pressure. Notice the absence of pressure term in the Poisson's equation.

We have checked all the equations in Einstein's gravity except for equation (\ref{pressure-relation1}). Using equations (\ref{pressure-relation2})-(\ref{pressure-relation4}), (\ref{E-conserv-I-pressure-N}) and (\ref{dot-kappa-pressure}) we can show that the left-hand-side of equation (\ref{pressure-relation1}) gives
\bea
   & & {4 \pi G \over c^2} a \Delta^{-1}  \nabla \cdot
       \left[ \left( \widetilde \varrho
       + {\widetilde p \over c^2} \right) {\bf v} \right]
   \nonumber \\
   & & \qquad
       - {8 \pi G \over c^2} a \Delta^{-1} \sum_J \left( {\bf v}_J \cdot \nabla
       {\widetilde p_J \over c^2} \right),
   \label{discrepancy}
\eea
where the second term is missing in the right-hand-side of the same equation; the form of missing term varies depending on the way of derivation. In Hwang \& Noh (2013c) we have attributed this failure as due to the higher order (in $c^{-2}$) nature of equation (\ref{pressure-relation1}). Accepting this argument we have checked the consistency of all the Einstein's equations.

For the collective fluid we can show
\bea
   & & \dot {\widetilde \varrho}
       + 3 {\dot a \over a} \left( \widetilde \varrho
       + {\widetilde p \over c^2} \right)
       + {1 \over a} \nabla \cdot \left[ \left( \widetilde \varrho
       + {\widetilde p \over c^2} \right) {\bf v}_I \right]
   \nonumber \\
   & & \qquad
       = {2 \over a} \sum_J {\bf v}_J \cdot \nabla {\widetilde p_J \over c^2},
   \label{continuity-p} \\
   & & \dot {\bf v}
       + {\dot a \over a} {\bf v}
       + {1 \over a} {\bf v} \cdot \nabla {\bf v}
       - {1 \over a} \nabla U
   \nonumber \\
   & & \qquad
       + {1 \over \widetilde \varrho + \widetilde p/ c^2}
       \left( {1 \over a} \nabla \widetilde p
       + {1 \over a} \nabla_j \Pi^{ij} \right)
   \nonumber \\
   & & \qquad
       = {1 \over \widetilde \varrho + \widetilde p/ c^2}
       \left[
       - {\bf v}_I {\dot {\widetilde p} \over c^2}
       + {2 \over a} \sum_J \left( {\bf v}_J - {\bf v} \right)
       {\bf v}_J \cdot \nabla {\widetilde p_J \over c^2}
       \right],
   \nonumber \\
   \label{Euler-p}
\eea
with
\bea
   & & \widetilde \varrho = \sum_K \widetilde \varrho_K, \quad
       \widetilde p = \sum_K \widetilde p_K,
   \nonumber \\
   & &
       \left( \widetilde \varrho
       + {\widetilde p \over c^2} \right) {\bf v}
       = \sum_K \left( \widetilde \varrho_K
       + {\widetilde p_K \over c^2} \right) {\bf v}_K,
   \nonumber \\
   & &
       \Pi_{ij}
       = - \left( \widetilde \varrho
       + {\widetilde p \over c^2} \right) v_i v_j
       + \sum_K \left( \widetilde \varrho_K
       + {\widetilde p_K \over c^2} \right) v_{Ki} v_{Kj}.
   \nonumber \\
   \label{fluid-p}
\eea
Terms in the right-hand-sides of equations (\ref{continuity-p}) and (\ref{Euler-p}) are the non-trivial contributions from relativistic pressure. In the absence of gravity equations (\ref{E-conserv-I-pressure-N}) and (\ref{mom-conserv-I-pressure-N}) or equations (\ref{continuity-p}) and (\ref{Euler-p}) have the proper special relativistic limits studied in equation (2.10.16) of Weinberg (1972) and equations (2.3) and (2.4) of Peacock (1999).

\section{Non relativistic pressure Fluids in the CDM-comoving gauge}

We consider the multi-component fluids with negligible pressure. By this, we ignore $\widetilde p_I/\widetilde \mu_I$ but keep $c^2 \widetilde p_{I,i}$ term in the momentum conservation equation. We consider the scalar-type perturbation in a flat background and ignore interaction terms among fluids. As one of the components we consider the cold-dark-matter (CDM), thus set $\widetilde p_c \equiv 0$. We take the CDM-comoving gauge by setting $\widehat v_c \equiv 0$ as the temporal gauge condition. The momentum conservation equation in equation (\ref{eq7-cov}) for the CDM component gives
\bea
   & & {\cal N} = 1.
\eea
Equation (\ref{eq6-cov}) for $I = c$, equation (\ref{eq4}), and equations (\ref{eq6-cov}) and (\ref{eq7-cov}) for other $I$ components give
\begin{widetext}
\bea
   & & \dot {\widetilde \varrho}_c
       + \widetilde \varrho_c \left( 3 H - \kappa \right)
       + {c \chi^{,i} \over a^2 (1 + 2 \varphi)} \widetilde \varrho_{c,i} = 0,
   \label{dot-varrho-c-CCG} \\
   & & \dot \kappa + 2 H \kappa
       - 4 \pi G \delta \varrho
       + {c \chi^{,i} \over a^2 (1 + 2 \varphi)} \kappa_{,i}
       = {1 \over 3} \kappa^2
       + 8 \pi G \widetilde \varrho \left( \widehat \gamma^2 - 1 \right)
       + c^2 \overline K^i_j \overline K^j_i
       + {1 \over 1 + 2 \varphi}
       {4 \pi G \over c^2} \left( \Pi^i_i
       - {1 \over 1 + 2 \varphi} \Pi_{ij}
       {\widehat v^i \widehat v^j \over c^2} \right),
   \nonumber \\
   \\
   & & \left[ {\partial \over \partial t}
       + {1 \over a (1 + 2 \varphi)}
       \left( \widehat v_I^i + {c \over a} \chi^{,i} \right)
       \nabla_i \right] \widetilde \varrho_I
   \nonumber \\
   & & \qquad
       + \widetilde \varrho_I
       \left\{ 3 H - \kappa
       + { \widehat v^i_{I,i} \over a ( 1 + 2 \varphi)}
       +{ \widehat v_I^i \varphi_{,i} \over a ( 1 + 2 \varphi )^2}
       + {1 \over \widehat \gamma_I}
       \left[ {\partial \over \partial t}
       + {1 \over a ( 1 + 2 \varphi )}
       \left( \widehat v_I^i + {c \over a} \chi^{,i} \right)
       \nabla_i \right] \widehat \gamma_I
       \right\}
       = 0,
   \\
   & & {\partial \over \partial t}
       \left( a \widehat \gamma_I \widehat v_{Ii} \right)
       + {1 \over 1 + 2 \varphi}
       \left( \widehat v_I^k + {c \over a} \chi^{,k} \right)
       \nabla_k \left( \widehat \gamma_I \widehat v_{Ii} \right)
       + {1 \over \widehat \gamma_I \widetilde \varrho_I} \widetilde p_{I,i}
       + {1 - \widehat \gamma_I^2 \over \widehat \gamma_I}
       {c^2 \varphi_{,i} \over 1 + 2 \varphi}
       + {c \over a} \widehat \gamma_I \widehat v_I^k
       \nabla_i \left( {\chi_{,k} \over 1 + 2 \varphi} \right)
       = 0.
   \label{dot-v-I-CCG}
\eea
Instead of equation for $\dot {\widehat v}_c$, we have equation of $\dot \kappa$. Together with equations (\ref{eq1})-(\ref{eq3}) and (\ref{eq5}), and the collective fluid quantities given in equation (\ref{fluid-Multi-2}) we have the complete set of equations.

\subsection{Newtonian limit in the general scale}

We further assume the weak gravity ($\alpha \ll 1$ and $\varphi \ll 1$), and the slow-motion ($\widehat v_I^i \widehat v_{Ii} / c^2 \ll 1$). Compared with the infinite-speed-of-light limit, here we do {\it not} assume the small-scale [$a^2 H^2 / (c^2 k^2) \ll 1$] condition. Equation (\ref{eq3}) gives
\bea
   & & {2 \over 3} \kappa_{,i}
       + {2 \over 3} {c \Delta \over a^2} \chi_{,i}
       = - {8 \pi G \over c^2} a \widetilde \varrho \widehat v_i,
   \label{eq3-CCG-N}
\eea
where we have properly considered the anisotropic stress term using equation (\ref{fluid-Multi-2}).

By identifying
\bea
   & & \kappa = - c {\Delta \over a^2} \chi
       \equiv - {1 \over a} \nabla \cdot {\bf v}_c
       \equiv - {\Delta \over a} v_c,
   \label{identification-CCG}
\eea
thus $c \chi /a = v_c$, equation (\ref{eq3-CCG-N}) is satisfied. Although we have introduced a new velocity component ${\bf v}_c (\equiv \nabla v_c)$, it does not necessarily imply that ${\bf v}_c$ is the velocity of the CDM component; the situation is not entirely clear in our multiple component situation, but see equations (\ref{dot-delta-c-CCG-N}) and (\ref{dot-v-c-CCG-N}) below; we still have $\widehat v_{ci} (\equiv - \widehat v_{c,i}) = 0$ as our CDM-comoving gauge condition. With this identification with ${\bf v}_I = v_I^i \equiv \widehat v_I^i$, equations (\ref{dot-varrho-c-CCG})-(\ref{dot-v-I-CCG}) become
\bea
   & & \dot {\widetilde \varrho}_c
       + 3 {\dot a \over a} \widetilde \varrho_c
       + {1 \over a} \nabla \cdot \left( \widetilde \varrho_c {\bf v}_c \right) = 0,
   \label{dot-delta-c-CCG-N} \\
   & & {1 \over a} \nabla \cdot \left( \dot {\bf v}_c
       + {\dot a \over a} {\bf v}_c \right)
       + 4 \pi G \delta \varrho
       + {1 \over a^2} \nabla \cdot \left( {\bf v}_c \cdot \nabla {\bf v}_c \right)
       = 0,
   \label{dot-v-c-CCG-N} \\
   & & \dot {\widetilde \varrho}_I
       + 3 {\dot a \over a} \widetilde \varrho_I
       + {1 \over a} \nabla \cdot \left[ \widetilde \varrho_I
       \left( {\bf v}_I + {\bf v}_c \right) \right] = 0,
   \label{dot-delta-I-CCG-N} \\
   & & \dot {\bf v}_I
       + {\dot a \over a} {\bf v}_I
       + {1 \over a} {\bf v}_I \cdot \nabla {\bf v}_I
       + {1 \over a} \left(
       {\bf v}_c \cdot \nabla {\bf v}_I
       + {\bf v}_I \cdot \nabla {\bf v}_c \right)
       + {1 \over a \widetilde \varrho_I} \nabla \widetilde p_I = 0,
   \label{dot-v-I-CCG-N}
\eea
where $\delta \varrho = \sum_J \delta \varrho_J$ from equation (\ref{fluid-Multi-2}).
Compared with the equations (\ref{mass-conserv-N-I})-(\ref{Poisson-eq-N}) in the Newtonian limit ($c \rightarrow \infty$ limit) in the zero-shear gauge and the uniform-expansion gauge, we have the additional presence of ${\bf v}_c$ contributions in equations (\ref{dot-delta-I-CCG-N}) and (\ref{dot-v-I-CCG-N}), the potential is missing in equation (\ref{dot-v-I-CCG-N}), and the Poisson's equation is absorbed equation (\ref{dot-v-c-CCG-N}). For the remaining Einstein equations, we can show that equation (\ref{eq1}) simply corresponds to equation (\ref{identification-CCG}), and both equation (\ref{eq2}) and equation (\ref{eq5}) together with equation (\ref{dot-v-c-CCG-N}) give
\bea
   & & c^2 {\Delta \over a^2} \varphi
       = - 4 \pi G \delta \varrho
       + {\dot a \over a} {1 \over a} \nabla \cdot {\bf v}_c
       + {1 \over 4 a^2} \left[
       \left( \nabla \cdot {\bf v}_c \right)^2
       - v_c^{\; ,ij} v_{c,ij} \right],
\eea which can be considered as a relation determining the curvature perturbation $\varphi$ from the fluid quantities. Notice that we have not assumed the small-scale limit. Thus the above results are valid in general scales including the super-horizon scale, whereas the proper Newtonian limit in the other two gauge conditions in Sec.\ \ref{sec:Newtonian} is valid only far inside the horizon.

The single component case was studied in Hwang, Noh \& Park (2014). In that case the above results remain valid by keeping only the $c$-component. In that case we further have not assumed the slow-motion limit ($|{\bf v}_c|^2/c^2 \ll 1$). In Hwang, Noh \& Park (2014) we have considered a fluid with non-relativistic pressure. In that case, equation (\ref{dot-v-c-CCG-N}) is replaced by
\bea
   & & {1 \over a} \nabla \cdot \left( \dot {\bf v}
       + {\dot a \over a} {\bf v} \right)
       + 4 \pi G \delta \varrho
       + {1 \over a^2} \nabla \cdot \left( {\bf v} \cdot \nabla {\bf v} \right)
       + {1 \over a^2} \nabla \cdot \left( {\nabla \widetilde p \over \widetilde \varrho} \right)
       = 0.
\eea

\subsection{Third order perturbations}

To the third order perturbations, equations (\ref{dot-varrho-c-CCG})-(\ref{dot-v-I-CCG}) become
\bea
   & & \delta \dot \varrho_c
       + 3 {\dot a \over a} \delta \varrho_c
       + {c \over a^2} \left( 1 - 2 \varphi \right)
       \chi^{,i} \delta \varrho_{c,i}
       - \left( \varrho_c + \delta \varrho_c \right) \kappa = 0,
   \label{dot-varrho-c-third-CCG} \\
   & & \dot \kappa + 2 {\dot a \over a} \kappa
       + {c \over a^2} \left( 1 - 2 \varphi \right) \chi^{,i} \kappa_{,i}
       - 4 \pi G \delta \varrho
       = {1 \over 3} \kappa^2
       + 8 \pi G \widetilde \varrho \left( \widehat \gamma^2 - 1 \right)
       + c^2 \overline K^i_j \overline K^j_i,
   \\
   & & \delta \dot \varrho_I
       + 3 {\dot a \over a} \delta \varrho_I
       - \varrho_I \left( 1 + \delta_I \right) \kappa
       + {1 \over a} \varrho_I \left( 1 + \delta_I - 2 \varphi
       - 2 \delta_I \varphi + 4 \varphi^2 \right) \widehat v_{I,i}^i
   \nonumber \\
   & & \qquad
       = - {1 \over a} \left( 1 - 2 \varphi \right)
       \left( \widehat v_I^i + {c \over a} \chi^{,i} \right) \delta \varrho_{I,i}
       + {1 \over c^2} \left( 1 - 2 \varphi \right) \widehat v_I^i \delta p_{I,i}
       + {1 \over c^2} {\dot a \over a} \varrho_I \left( 1 + \delta_I - 2 \varphi \right)
       \widehat v_I^i \widehat v_{Ii}
   \nonumber \\
   & & \qquad
       + {1 \over c a^2} \varrho_I \widehat v_I^i
       \widehat v_I^k \chi_{,ik}
       + {1 \over c^2} \varrho_I \dot \varphi
       \widehat v_I^i \widehat v_{Ii}
       - {1 \over a} \varrho_I
       \left( 1 + \delta_I - 4 \varphi \right) \widehat v_I^i \varphi_{,i},
   \\
   & & \dot {\widehat v}_{Ii}
       + {\dot a \over a} \widehat v_{Ii}
       + \left( 1 - \delta_I + \delta_I^2 \right)
       {\delta p_{I,i} \over a \varrho_I}
   \nonumber \\
   & & \qquad
       = - {1 \over a} \left[ \left( 1 - 2 \varphi \right) \widehat v_I^k
       \left( {1 \over 2} \widehat v_{Ik} + {c \over a} \chi_{,k} \right) \right]_{,i}
       + {1 \over c^2} \widehat v_I^k \widehat v_{Ik}
       \left( {\dot a \over a} \widehat v_{Ii}
       + {3 \over 2} {\delta p_{I,i} \over a \varrho_I} \right)
       + {1 \over c^2} \widehat v_{Ii} \widehat v_I^k
       {\delta p_{I,k} \over a \varrho_I},
   \label{dot-v-I-third-CCG}
\eea
where
\bea
   & & \delta \varrho
       = \sum_J \Bigg\{
       \delta \varrho_J
       + {1 \over c^2} \varrho_J
       \Bigg[ \left( 1 + \delta_J - 2 \varphi \right) \widehat v_J^i \widehat v_{Ji}
       + {2 \over \sum_L \varrho_L}
       \left( \sum_K \delta \varrho_K \widehat v_K^i \right)
       \left( {1 \over \sum_L \varrho_L} \sum_M \varrho_M \widehat v_{Mi}
       - \widehat v_{Ji} \right)
   \nonumber \\
   & & \qquad
       + {1 \over \sum_L \varrho_L}
       \left( 1 + \delta_J - 2 \varphi
       - {\sum_M \delta \varrho_M \over \sum_L \varrho_L} \right)
       \left( \sum_K \varrho_K \widehat v_K^i \right)
       \left( {1 \over \sum_L \varrho_L} \sum_M \varrho_M \widehat v_{Mi}
       - 2 \widehat v_{Ji} \right)
       \Bigg] \Bigg\},
   \nonumber \\
   & & \widehat \gamma^2 - 1
       = {1 \over c^2} {1 \over \left( \sum_L \varrho_L \right)^2}
       \left( \sum_J \varrho_J \widehat v_J^i \right)
       \left[
       \left( 1 - 2 \varphi
       - 2 {\sum_M \delta \varrho_M \over \sum_L \varrho_L} \right)
       \left( \sum_K \varrho_K \widehat v_{Ki} \right)
       - 2 \sum_K \delta \varrho_K \widehat v_{Ki} \right],
   \nonumber \\
   & & c^2 \overline K^i_j \overline K^j_i
       = {c^2 \over a^4} \left( 1 - 4 \varphi \right)
       \left[ \chi^{,ij} \chi_{,ij}
       - {1 \over 3} \left( \Delta \chi \right)^2 \right].
\eea We have used $\widehat v_i = - \widehat v_{,i}$ for the scalar-type perturbation. As we took the CDM comoving gauge, we have $\widehat v_{Ji} = 0$ for $J = c$.

From the above set of equations, we need relations determining $\chi$ and $\varphi$ to the second order perturbation; in the single component case we needed $\varphi$ only to the linear order (Hwang \& Noh 2005). From equations (\ref{eq3}) and (\ref{eq2}), respectively, we have \bea
   & & c {\Delta \over a^2} \chi_{,i}
       = - \left( 1 + 2 \varphi \right)
       \left( \kappa_{,i}
       + {12 \pi G \over c^2} a \widetilde \varrho \widehat v_i \right)
       - {c \over a^2} \left(
       \varphi_{,i} \Delta \chi
       + {3 \over 2} \chi_{,i} \Delta \varphi
       + 2 \varphi^{,j} \chi_{,ij}
       + {1 \over 2} \chi^{,j} \varphi_{,ij} \right),
   \\
   & & c^2 {\Delta \over a^2} \varphi
       = - \left( 1 + 4 \varphi \right)
       \left( 4 \pi G \delta \varrho
       + {\dot a \over a} \kappa \right)
       + {1 \over 6} \kappa^2
       - 4 \pi G \varrho \left( \widehat \gamma^2 - 1 \right)
       + {3 \over 2} {c^2 \over a^2} \varphi^{,i} \varphi_{,i}
       - {c^2 \over 4} \overline K^i_j \overline K^j_i,
\eea
to the second order perturbation, and from equation (\ref{eq1}), we have
\bea
   & & \dot \varphi
       = - {1 \over 3} \left( \kappa
       + c {\Delta \over a^2} \chi \right),
   \label{dot-phi-p}
\eea
to the linear order.
\end{widetext}

\section{Minimally coupled scalar fields}
                                            \label{sec:fields}

We consider an action given as (Hwang \& Noh 2000)
\bea
   & & S = \int d^4 x \sqrt{- \widetilde g} \left[
       {1 \over 16 \pi G} \widetilde R
       - {1 \over 2} \widetilde \phi^{I; c} \widetilde \phi_{I,c}
       - \widetilde V ( \widetilde \phi^J ) \right],
   \nonumber \\
\eea
where $\widetilde \phi_I$ are minimally coupled scalar fields with $I, J \dots = 1, 2, \dots N$. In this section, except for Sec.\ \ref{sec:MSF-fluid}, we assume the summation convention for the repeated indices of $I, J, \dots$, and set $c \equiv 1 \equiv \hbar$ in the presence of the scalar field. The energy-momentum tensor becomes
\bea
   & & \widetilde T_{ab}
       = \widetilde \phi^I_{\;,a} \widetilde \phi_{I,b}
       - \left( {1 \over 2} \widetilde \phi^{I;c} \widetilde \phi_{I,c}
       + \widetilde V \right) \widetilde g_{ab}.
   \label{Tab-MSFs}
\eea
The equation of motion for each field is
\bea
   & & \widetilde \phi^{\;\; ;c}_{I\;\;\; c}
       = \widetilde V_{,I},
   \label{EOM-MSFs}
\eea
where $\widetilde V_{,I} \equiv \partial \widetilde V / (\partial \widetilde \phi^I)$.

As in the multi-component fluids, equations (\ref{eq1})-(\ref{eq5}) remain valid as the fluid quantities considered as the collective ones. Thus we need the collective fluid quantities expressed in terms of individual field quantities. Instead of the (energy and momentum) conservation equations we have the equation of motion of each component.

\subsection{Collective fluid quantities}

From equations (\ref{Tab-MSFs}) and (\ref{Tab-decomposition}) we have
\bea
   & & \widetilde \mu = {1 \over 2}
       \widetilde {\dot {\widetilde \phi}}^I
       \widetilde {\dot {\widetilde \phi}}_I
       + {1 \over 2} \widetilde h^{cd}
       \widetilde \phi^I_{\;,c}
       \widetilde \phi_{I,d}
       + \widetilde V,
   \nonumber \\
   & &
       \widetilde p = {1 \over 2}
       \widetilde {\dot {\widetilde \phi}}^I
       \widetilde {\dot {\widetilde \phi}}_I
       - {1 \over 6} \widetilde h^{cd}
       \widetilde \phi^I_{\;,c}
       \widetilde \phi_{I,d}
       - \widetilde V, \quad
       \widetilde q_a =
       - \widetilde {\dot {\widetilde \phi}}^I
       \widetilde h^b_a \widetilde \phi_{I,b},
   \nonumber \\
   & &
       \widetilde \pi_{ab}
       = \left( \widetilde h^c_a \widetilde h^d_b
       - {1 \over 3} \widetilde h_{ab} \widetilde h^{cd} \right)
       \widetilde \phi^I_{\;,c} \widetilde \phi_{I,d},
   \label{fluid-MSFs}
\eea
where $\widetilde {\dot {\widetilde \phi}}_I
\equiv \widetilde \phi_{I,c} \widetilde u^c$.

As we take the energy-frame ($\widetilde q_a \equiv 0$) we have $\widetilde {\dot {\widetilde \phi}}^I \widetilde h^b_a \widetilde \phi_{I,b} = 0$. Using equation (\ref{fluid-PT}) we have \bea
   & & \widehat v_i
       = - {1 \over a \widehat \gamma}
       { \widetilde {\dot {\widetilde \phi}}^I \widetilde \phi_{I,i}
       \over
       \widetilde {\dot {\widetilde \phi}}^J
       \widetilde {\dot {\widetilde \phi}}_J}.
   \label{v-MSFs}
\eea We can show \bea
   & & {\widetilde {\dot {\widetilde \phi}}}_I
       = \widetilde \phi_{I,c} \widetilde u^c
       = \widehat \gamma \left(
       {D \over D t}
       + {\widehat v^i \over a ( 1 + 2 \varphi )} \nabla_i \right)
       \widetilde \phi_I,
\eea
where we introduced \bea
   & & {D \over Dt}
       \equiv {1 \over N} \left( \partial_0
       - N^i \nabla_i \right)
       = {1 \over {\cal N}} \left( {\partial \over \partial t}
       + {\chi^i \over a^2 (1 + 2 \varphi)} \nabla_i \right).
   \nonumber \\
\eea

\begin{widetext}
The collective fluid quantities in equations (\ref{fluid-MSFs}) and (\ref{v-MSFs}) become \bea
   & & \widehat v_i
       = - {1 \over a \widehat \gamma^2}
       { \left[ \left( {D \over D t}
       + {\widehat v^j \over a ( 1 + 2 \varphi )} \nabla_j \right)
       \widetilde \phi^I \right] \widetilde \phi_{I,i}
       \over
       \left[ \left( {D \over D t}
       + {\widehat v^k \over a ( 1 + 2 \varphi )} \nabla_k \right)
       \widetilde \phi^J \right]
       \left[ \left( {D \over D t}
       + {\widehat v^\ell \over a ( 1 + 2 \varphi )} \nabla_\ell \right)
       \widetilde \phi_J \right]},
   \nonumber \\
   & & \widetilde \mu
       = \widehat \gamma^2
       \left[ \left( {D \over D t}
       + {\widehat v^i \over a ( 1 + 2 \varphi )} \nabla_i \right)
       \widetilde \phi^I \right]
       \left[ \left( {D \over D t}
       + {\widehat v^j \over a ( 1 + 2 \varphi )} \nabla_j \right)
       \widetilde \phi_I \right]
       - {1 \over 2}
       {D \widetilde \phi^I \over Dt}
       {D \widetilde \phi_I \over Dt}
       + {1 \over 2} {\widetilde \phi^{I|i} \widetilde \phi_{I,i}
       \over a^2 ( 1 + 2 \varphi )}
       + \widehat V,
   \nonumber \\
   & & \widetilde p
       = {1 \over 3} \widehat \gamma^2
       \left[ \left( {D \over D t}
       + {\widehat v^i \over a ( 1 + 2 \varphi )} \nabla_i \right)
       \widetilde \phi^I \right]
       \left[ \left( {D \over D t}
       + {\widehat v^j \over a ( 1 + 2 \varphi )} \nabla_j \right)
       \widetilde \phi_I \right]
       + {1 \over 6}
       {D \widetilde \phi^I \over Dt}
       {D \widetilde \phi_I \over Dt}
       - {1 \over 6} {\widetilde \phi^{I|i} \widetilde \phi_{I,i}
       \over a^2 ( 1 + 2 \varphi )}
       - \widehat V,
   \nonumber \\
   & & \Pi_{ij}
       =
       \left\{ \widehat \gamma^2
       \left[ \left( {D \over D t}
       + {\widehat v^k \over a ( 1 + 2 \varphi )} \nabla_k \right)
       \widetilde \phi^I \right] \widehat v_i
       + {1 \over a} \widetilde \phi^I_{\;,i} \right\}
       \left\{ \widehat \gamma^2
       \left[ \left( {D \over D t}
       + {\widehat v^\ell \over a ( 1 + 2 \varphi )} \nabla_\ell \right)
       \widetilde \phi_I \right] \widehat v_j
       + {1 \over a} \widetilde \phi_{I,j} \right\}
   \nonumber \\
   & & \qquad
       - {1 \over 3} \left[ \left( 1 + 2 \varphi \right) \gamma_{ij}
       + \widehat \gamma^2 \widehat v_i \widehat v_j \right]
       \Bigg\{ \widehat \gamma^2
       \left[ \left( {D \over D t}
       + {\widehat v^k \over a ( 1 + 2 \varphi )} \nabla_k \right)
       \widetilde \phi^I \right]
       \left[ \left( {D \over D t}
       + {\widehat v^\ell \over a ( 1 + 2 \varphi )} \nabla_\ell \right)
       \widetilde \phi_I \right]
   \nonumber \\
   & & \qquad
       - {D \widetilde \phi^I \over Dt}
       {D \widetilde \phi_I \over Dt}
       + {\widetilde \phi^{I|k} \widetilde \phi_{I,k}
       \over a^2 ( 1 + 2 \varphi )}
       \Bigg\}.
   \label{fluid-MSFs-exact}
\eea
As the right-hand-side of $\widehat v_i$ in equation (\ref{fluid-MSFs-exact}) contains $\widehat v_k$, in order to express $\widehat v_i$ purely in terms of individual field quantities, we have to solve the relation perturbatively by iteration. The other collective fluid quantities also contain $\widehat v_i$ which should be replaced by thus obtained individual ones.

For example, to the linear and second order, we have
\bea
   & & \widehat v_i
       = - {1 \over a}
       { \dot \phi^I \widetilde \phi_{I,i}
       \over
       \dot \phi^J \dot \phi_J}, \quad
       \widehat v_i
       = - {{\cal N} \over a}
       { \dot {\widetilde \phi}^I \widetilde \phi_{I,i}
       \over
       \dot {\widetilde \phi}^J \dot {\widetilde \phi}_J},
   \label{v-MSFs-second-order}
\eea respectively, and to the third order, we have \bea
   & & \widehat v_i
       = - {1 \over a} \left( 1 - \widehat v^m \widehat v_m \right)
       { \left[ \left( {D \over D t}
       + {1 \over a} \widehat v^j \nabla_j \right)
       \widetilde \phi^I \right] \widetilde \phi_{I,i}
       \over
       \left[ \left( {D \over D t}
       + {1 \over a} \widehat v^k \nabla_k \right)
       \widetilde \phi^J \right]
       \left[ \left( {D \over D t}
       + {1 \over a}\widehat v^\ell \nabla_\ell \right)
       \widetilde \phi_J \right]},
   \label{v-MSFs-third-order}
\eea
where we can replace $\widehat v^k$'s in the right-hand-side by $\widehat v^k$ to the linear order in equation (\ref{v-MSFs-second-order}). This can be continued to any higher order perturbation.

\subsection{Equations of motion}

The equation of motion in equation (\ref{EOM-MSFs}) gives
\bea
   & & - \widetilde \phi^{\;\; ;c}_{I\;\;\; c}
       = {D^2 \widetilde \phi_I \over Dt^2}
       + \left( 3 H
       - \kappa \right)
       {D \widetilde \phi_I \over Dt}
       - {( {\cal N} \sqrt{1 + 2 \varphi}
       \widetilde \phi_I^{\;\;|i} )_{|i} \over a^2 {\cal N} (1 + 2 \varphi)^{3/2}}
   \nonumber \\
   & & \qquad
      = {1 \over {\cal N}^2} \Bigg\{
       \ddot {\widetilde \phi}_I
       + \left( 3 H {\cal N}
       - {\cal N} \kappa
       - {\dot {\cal N} \over {\cal N}}
       - {\chi^i {\cal N}_{,i} \over
       a^2 {\cal N} ( 1 + 2 \varphi )} \right) \dot {\widetilde \phi}_I
       + {2 \chi^i \over
       a^2 ( 1 + 2 \varphi )} \dot {\widetilde \phi}_{I,i}
   \nonumber \\
   & & \qquad
       - {1 \over a^2 (1 + 2 \varphi)}
       \left( {\cal N}^2 \gamma^{ij}
       - {\chi^i \chi^j \over a^2 (1 + 2 \varphi)} \right) \widetilde \phi_{I,i|j}
       + \Bigg[ - {{\cal N}^2 \over a^2 (1 + 2 \varphi)}
       \left( {{\cal N}^{|i} \over {\cal N}}
       + {\varphi^{|i} \over 1 + 2 \varphi} \right)
   \nonumber \\
   & & \qquad
       + \left( 3 H {\cal N}
       - {\cal N} \kappa
       - {\dot {\cal N} \over {\cal N}}
       - {\chi^k {\cal N}_{,k} \over
       a^2 {\cal N} ( 1 + 2 \varphi )} \right)
       {\chi^i \over a^2 (1 + 2 \varphi)}
       + \left( {\chi^i \over a^2 (1 + 2 \varphi)}
       \right)^{\displaystyle\cdot}
       + {\chi^k \over a^4 (1 + 2 \varphi)}
       \left( {\chi^i \over 1 + 2 \varphi} \right)_{|k}
       \Bigg] \widetilde \phi_{I,i}
       \Bigg\}
   \nonumber \\
   & & \qquad
       =  - \widetilde V_{,I}.
   \label{EOMs-perturbed}
\eea
\end{widetext}

Equations (\ref{eq1})-(\ref{eq5}) and (\ref{EOMs-perturbed}) together with the collective fluid quantities in equation (\ref{fluid-MSFs-exact}) provide a complete set of nonlinear and exact perturbation equations in the presence of the multiple fields. In the perturbation theory we may introduce \bea
   & & \widetilde \phi_I \equiv \phi_I + \delta \phi_I,
\eea
where $\phi_I$'s are the background scalar fields, and $\delta \phi_I$'s are functions of space-time with arbitrary amplitudes.

In addition to the fundamental gauge conditions in equation (\ref{temporal-gauges-NL}), in the multi-component field case, for a chosen $I$-component we have the following gauge condition available
\bea
   & & {\rm uniform-}I{\rm -component\!-\!field \; gauge:} \hskip .5cm \delta \phi_{I} = 0,
   \nonumber \\
   \label{temporal-gauges-I-MSFs}
\eea
to the fully nonlinear order. Under this gauge condition for a chosen $I$-component, equation (\ref{EOMs-perturbed}) gives
\bea
   & & \ddot {\phi}_I
       + \left( 3 H {\cal N}
       - {\cal N} \kappa
       - {\dot {\cal N} \over {\cal N}}
       - {\chi^i {\cal N}_{,i} \over
       a^2 {\cal N} ( 1 + 2 \varphi )} \right) \dot {\phi}_I
   \nonumber \\
   & & \qquad
       = - {\cal N}^2 V_{,I}.
\eea

\subsection{Fluid formulation}
                                       \label{sec:MSF-fluid}

For the scalar fields, instead of the equations of motion, we can also use the energy and momentum conservation equations of the individual component in the fluid formulation. In order for this, we need to introduce fluid quantities of the individual component of the field. In the following we do not assume the summation convention for $I, J, \dots$. We may introduce \bea
   & & \widetilde T_{Iab}
       = \widetilde \phi^I_{\;,a} \widetilde \phi_{I,b}
       - \left( {1 \over 2} \widetilde \phi^{I;c} \widetilde \phi_{I,c}
       + \widetilde V_I \right) \widetilde g_{ab},
   \label{Tab-MSFs-I}
\eea
with
\bea
   & & \widetilde T_{ab} = \sum_J \widetilde T_{Jab}, \quad
       \widetilde V \equiv \sum_J \widetilde V_J.
\eea
Decomposition of the potential into the individual ones $\widetilde V_I$ has ceratin degree of arbitrariness, but does not imply that we ignore the interaction among the fields.
From equations (\ref{Tab-MSFs-I}) and (\ref{Tab-decomposition-I}) we have \bea
   & &
       \widetilde \mu_I = {1 \over 2}
       \left( \widetilde \phi_{I,b} \widetilde u^{Ib} \right)^2
       + \widetilde V_I, \quad
       \widetilde p_I = {1 \over 2}
       \left( \widetilde \phi_{I,b} \widetilde u^{Ib} \right)^2
       - \widetilde V_I,
   \nonumber \\
   & &
       \widetilde u_{Ia} = - {\widetilde \phi_{I,a} \over
       \widetilde \phi_{I,b} \widetilde u^{Ib}}, \quad
       \widetilde \pi_{Iab} = 0,
   \label{fluid-MSFs-I}
\eea
where we have used the energy-frame condition $\widetilde q_{Ia} \equiv 0$.
Using the fluid quantities of individual field in equation (\ref{fluid-MSFs-I}) we can show that the collective fluid quantities in equation (\ref{fluid-MSFs}) can be identified with the ones in the fluid formulation in equation (\ref{fluid-Multi-1}).
Therefore, instead of the equations of motion in equation (\ref{EOMs-perturbed}) we can use the energy and momentum conservation equations in equations (\ref{eq6-ADM})-(\ref{eq7-cov}) using the individual fluid quantities identified in equation (\ref{fluid-MSFs-I}).

The interaction terms can be read from
\bea
   & & \widetilde T_{Ia;b}^{\;\;b}
       = {\partial \widetilde V \over \partial \widetilde \phi^I}
       \widetilde \phi^I_{,a}
       - \sum_K {\partial \widetilde V_I \over \partial \widetilde \phi^K}
       \widetilde \phi^K_{,a}
       \equiv \widetilde I_{Ia},
   \label{interaction-MSF}
\eea
where we have used the equations of motion in equation (\ref{EOM-MSFs}). For $\widetilde V_I = \widetilde V_I (\widetilde \phi^I)$ for all $I$'s, we have $\widetilde I_{Ia} = 0$.

From equation (\ref{fluid-MSFs-I}) we have fluid quantities of individual component
\bea
   & &
       \widetilde \mu_I
       = {1 \over 2} \widehat \gamma_I^2
       \left[ \left( {D \over Dt} + {\widehat v_I^k \over a (1 + 2 \varphi)} \nabla_k \right) \widetilde \phi_I \right]^2 + \widetilde V_I,
   \nonumber \\
   & &
       \widetilde p_I
       = {1 \over 2} \widehat \gamma_I^2
       \left[ \left( {D \over Dt} + {\widehat v_I^k \over a (1 + 2 \varphi)} \nabla_k \right) \widetilde \phi_I \right]^2 - \widetilde V_I,
   \nonumber \\
   & & \widehat v_{Ii}
       = - {\widetilde \phi_{I,i} \over a \widehat \gamma_I^2
       \left( {D \over Dt}
       + {\widehat v_I^k \over a (1 + 2 \varphi)} \nabla_k \right) \widetilde \phi_I},
   \nonumber \\
   & & \Pi_{Iij} = 0.
   \label{fluids-MSF-I}
\eea
Using these relations we can derive equation (\ref{fluid-MSFs-exact}) from equation (\ref{fluid-Multi-2}). In order to show this we need the following two relations
\bea
   & & \left( {D \over Dt} + {\widehat v^k \over a (1 + 2 \varphi)}
       \nabla_k \right) \widetilde \phi_I
   \nonumber \\
   & & \qquad
       = \widehat \gamma_I^2
       \left( 1 - {\widehat v_\ell \widehat v_I^\ell
       \over 1 + 2 \varphi} \right)
       \left( {D \over Dt} + {\widehat v_I^k \over a (1 + 2 \varphi)} \nabla_k \right) \widetilde \phi_I,
   \nonumber \\
   & & \widehat \gamma_I^2
       \left[
       \left( {D \over Dt} + {\widehat v_I^k \over a (1 + 2 \varphi)} \nabla_k \right) \widetilde \phi_I \right]^2
   \nonumber \\
   & & \qquad
       = \left( {D \widetilde \phi_I \over Dt} \right)^2
       - {\widetilde \phi_I^{\;\;|i} \widetilde \phi_{I,i}
       \over a^2 ( 1 + 2 \varphi )},
\eea
which follow from identities
\bea
   & & \widetilde \phi_{I,a} \widetilde u^a
       = - \widetilde \phi_{I,b} \widetilde u_I^b
       \widetilde u_{Ia} \widetilde u^a, \quad
       \left( \widetilde \phi_{I,a} \widetilde u_I^a \right)^2
       = - \widetilde \phi_{I,a} \widetilde \phi_I^{\; ;a},
   \nonumber \\
\eea
respectively.

\section{Linear order}

We consider linear perturbations in the Friedmann background world model. For generality we recover the anisotropic stress of individual component; to the linear order, it appears only in the tracefree ADM propagation and the momentum conservation equations.
To the background order, equations (\ref{eq2}), (\ref{eq4}), (\ref{eq6-ADM}) and (\ref{eq6-cov}) give \bea
   & & H^2 = {8 \pi G \over 3 c^2} \mu
       - {\overline K c^2 \over a^2}
       + {\Lambda c^2 \over 3},
   \nonumber \\
   & &
       \dot H + H^2 = - {4 \pi G \over 3 c^2} \left( \mu + 3 p \right)
       + {\Lambda c^2 \over 3},
   \nonumber \\
   & &
       \dot \mu_I + 3 H \left( \mu_I + p_I \right)
       = - {c \over a} I_{I0}.
\eea
To the linear order in perturbation, equations (\ref{eq1})-(\ref{eq5}) and (\ref{eq6-ADM})-(\ref{eq7-cov}) give (Bardeen 1988, Hwang 1991, Hwang \& Noh 2013a)
\bea
   & & \kappa = 3 H \alpha - 3 \dot \varphi
       - c {\Delta \over a^2} \chi,
   \label{eq1-linear} \\
   & & {4 \pi G \over c^2} \delta \mu
       + H \kappa
       + c^2 {\Delta + 3 \overline K \over a^2} \varphi = 0,
   \\
   & & \kappa + c {\Delta + 3 \overline K \over a^2} \chi
       - {12 \pi G \over c^4} a \left( \mu + p \right) \widehat v = 0,
   \\
   & & \dot \kappa + 2 H \kappa
       + \left( 3 \dot H + c^2 {\Delta \over a^2} \right) \alpha
       = {4 \pi G \over c^2} \left( \delta \mu + 3 \delta p \right),
   \\
   & & \varphi + \alpha
       - {1 \over c} \left( \dot \chi + H \chi \right)
       = - {8 \pi G \over c^4} \Pi,
   \\
   & & \delta \dot \mu
       + 3 H \left( \delta \mu + \delta p \right)
       + \left( \mu + p \right) \left( 3 H \alpha
       - \kappa - {\Delta \over a} \widehat v \right)
       = 0,
   \nonumber \\
   \label{eq6-linear} \\
   & & {1 \over a^4} \left[ a^4 \left( \mu + p \right)
       \widehat v \right]^{\displaystyle\cdot}
       = {c^2 \over a} \left[ \delta p
       + \left( \mu + p \right) \alpha
       + {2 \over 3} {\Delta + 3 \overline K \over a^2} \Pi \right],
   \nonumber \\
   \label{eq7-linear} \\
   & & \delta \dot \mu_I
       + 3 H \left( \delta \mu_I + \delta p_I \right)
       + \left( \mu_I + p_I \right) \left( 3 H \alpha
       - \kappa - {\Delta \over a} \widehat v_I \right)
   \nonumber \\
   & & \qquad
       = - {c \over a} \delta I_{I0},
   \label{eq6-I-linear} \\
   & & {1 \over a^4} \left[ a^4 \left( \mu_I + p_I \right)
       \widehat v_I \right]^{\displaystyle\cdot}
   \nonumber \\
   & & \qquad
       = {c^2 \over a} \left[ \delta p_I
       + \left( \mu_I + p_I \right) \alpha
       + {2 \over 3} {\Delta + 3 \overline K \over a^2} \Pi_I
       - I_I \right],
   \nonumber \\
   \label{eq7-I-linear}
\eea
for the scalar-type perturbation.
Equations (\ref{eq3}), (\ref{eq5}) and (\ref{eq7-ADM}) give
\bea
   & & c {\Delta + 2 \overline K \over a^2} \chi^{(v)}_i
       = - {16 \pi G \over c^4} a \left( \mu + p \right) \widehat v^{(v)}_i,
   \\
   & & {c \over a} \left( \dot \chi^{(v)}_i
       + H \chi^{(v)}_i \right)
       = {8 \pi G \over c^2} \Pi^{(v)}_i,
   \\
   & & {1 \over a^4} \left[ a^4 \left( \mu + p \right)
       \widehat v_{i}^{(v)} \right]^{\displaystyle\cdot}
       = - c^2 {\Delta + 2 \overline K \over 2 a^2} \Pi^{(v)}_{i},
   \\
   & & {1 \over a^4} \left[ a^4 \left( \mu_I + p_I \right)
       \widehat v_{Ii}^{(v)} \right]^{\displaystyle\cdot}
       = - c^2 {\Delta + 2 \overline K \over 2 a^2} \Pi^{(v)}_{Ii}
       + {c^2 \over a} I_{Ii}^{(v)},
   \nonumber \\
\eea
for the vector-type perturbation. For fluid quantities equation (\ref{fluid-Multi-2}) gives
\bea
   & & \mu = \sum_J \mu_J, \quad
       p = \sum_J p_J,
\eea
to the background order, and
\bea
   & & \delta \mu = \sum_J \delta \mu_J, \quad
       \delta p = \sum_J \delta p_J,
   \nonumber \\
   & &
       \widehat v = {\sum_J ( \mu_J + p_J ) \widehat v_J \over
       \sum_K (\mu_K + p_K)}, \quad
       \widehat v^{(v)}_i = {\sum_J ( \mu_J + p_J )
       \widehat v^{(v)}_{Ji} \over
       \sum_K (\mu_K + p_K)},
   \nonumber \\
   & &
       \Pi = \sum_J \Pi_J, \quad
       \Pi^{(v)}_i = \sum_J \Pi^{(v)}_{Ji},
\eea
to the perturbed order.

\subsection{Scalar fields}

For the scalar fields, equations (\ref{EOMs-perturbed}) and (\ref{fluid-MSFs-exact}) give
\bea
   & & \ddot \phi_I + 3 H \dot \phi_I + V_{,I} = 0,
   \\
   & & \mu = {1 \over 2} \sum_J \dot \phi_J^2 + V, \quad
       p = {1 \over 2} \sum_J \dot \phi_J^2 - V,
\eea
to the background order, and
\bea
   & & \delta \ddot \phi_I
       + 3 H \delta \dot \phi_I
       - {\Delta \over a^2} \delta \phi_I
       + \sum_J V_{,IJ} \delta \phi_J
   \nonumber \\
   & & \qquad
       = 2 \ddot \phi_I \alpha
       + \dot \phi_I \left( \dot \alpha
       + 3 H \alpha + \kappa \right),
   \\
   & &
       \delta \mu
       = \sum_J \left(
       \dot \phi^J \delta \dot \phi_J
       - \dot \phi_J^2 \alpha
       + V_{,J} \delta \phi^J \right),
   \nonumber \\
   & &
       \delta p
       = \sum_J \left(
       \dot \phi^J \delta \dot \phi_J
       - \dot \phi_J^2 \alpha
       - V_{,J} \delta \phi^J \right),
   \nonumber \\
   & & \widehat v = {1 \over a}
       {\sum_J \dot \phi^J \delta \phi_J
       \over \sum_K \dot \phi_K^2}, \quad
       \widehat v^{(v)}_i = 0, \quad
       \Pi_{ij} = 0,
\eea
to the linear perturbation order. Using the individual components, equation (\ref{fluids-MSF-I}) give
\bea
   & & \mu_I = {1 \over 2} \dot \phi_I^2 + V_I, \quad
       p_I = {1 \over 2} \dot \phi_I^2 - V_I,
   \nonumber \\
   & & \delta \mu_I = \dot \phi_I \delta \dot \phi_I
       - \dot \phi_I^2 \alpha
       + \sum_J V_{I,J} \delta \phi^J,
   \nonumber \\
   & &
       \delta p_I = \dot \phi_I \delta \dot \phi_I
       - \dot \phi_I^2 \alpha
       - \sum_J V_{I,J} \delta \phi^J,
   \nonumber \\
   & &
       \widehat v_I = {\delta \phi_I \over a \dot \phi_I}, \quad
       \widehat v^{(v)}_i = 0, \quad
       \Pi_{ij} = 0.
\eea
The interaction term follows from equation (\ref{interaction-MSF}) as
\bea
   & & I_{I0} = a \left( {\partial V \over \partial \phi^I} \dot \phi^I
       - \sum_K {\partial V_I \over \partial \phi^K} \dot \phi^K \right),
   \\
   & & \delta I_{I0} = a \Bigg[ {\partial V \over \partial \phi^I}
       \delta \dot \phi^I
       + \sum_J {\partial^2 V \over \partial \phi^J \partial \phi^I}
       \dot \phi^I \delta \phi^J
   \nonumber \\
   & & \qquad
       - \sum_K \left( {\partial V_I \over \partial \phi^K}
       \delta \dot \phi^K
       + \sum_J {\partial^2 V_I \over \partial \phi^J \partial \phi^K}
       \dot \phi^K \delta \phi^J \right) \Bigg],
   \nonumber \\
   & & I_I = {\partial V \over \partial \phi^I}
       \delta \phi^I
       - \sum_K {\partial V_I \over \partial \phi^K}
       \delta \phi^K, \quad
       I^{(v)}_{Ii} = 0.
\eea
As we have shown below equation (\ref{interaction-MSF}), for $\widetilde V_I = \widetilde V_I (\widetilde \phi^I)$ for all $I$'s, we have $\widetilde I_{Ia} = 0$.

For fluids system, see Kodama \& Sasaki (1986, 1987). For minimally coupled scalar fields system, see Sasaki \& Stewart (1996) and Hwang \& Noh (2000). For fluids-field system, see Hwang \& Noh (2002).

\begin{widetext}
\subsection{Using relative variables}

Kodama \& Sasaki (1984) introduced the following relative variables
\bea
   & & S_{IJ} \equiv {\delta \mu_I \over \mu_I + p_I}
         - {\delta \mu_J \over \mu_J + p_J}, \quad
         \widehat v_{IJ} \equiv \widehat v_I - \widehat v_J, \quad
         e_{IJ} \equiv {e_I \over \mu_I + p_I}
         - {e_J \over \mu_J + p_J}, \quad
         \Pi_{IJ} \equiv {\Pi_{I} \over \mu_I + p_I}
         - {\Pi_{J} \over \mu_J + p_J},
   \nonumber \\
   & &
         \delta I_{IJ0} \equiv {\delta I_{I0} \over \mu_I + p_I}
         - {\delta I_{J0} \over \mu_J + p_J}, \quad
         I_{IJi} \equiv {I_{Ii} \over \mu_I + p_I}
         - {I_{Ji} \over \mu_J + p_J},
\eea
where
\bea
   & & e_I \equiv \delta p_I - c_I^2 \delta \mu_I, \quad
         c_I^2 \equiv {\dot p_I \over \dot \mu_I}.
\eea
From equations (\ref{eq6-I-linear}) and (\ref{eq7-I-linear}), respectively, we have (Kodama \& Sasaki 1984)
\bea
   & & \dot S_{IJ} - {\Delta \over a} \widehat v_{IJ}
         + 3 H e_{IJ}
         = - {c \over a} \delta I_{IJ0}
         + {c \over 2 a} \left( {1 + c_I^2 \over \mu_I + p_I} I_{I0}
         + {1 + c_J^2 \over \mu_J +  p_J} I_{J0} \right) S_{IJ}
   \nonumber \\
   & & \qquad
         + {c \over a} \left( {1 + c_I^2 \over \mu_I + p_I} I_{I0}
         - {1 + c_J^2 \over \mu_J +  p_J} I_{J0} \right)
         \left[ {1 \over 2} \sum_K {\mu_K + p_K \over \mu + p} \left( S_{IK} + S_{JK} \right)
         + {\delta \mu \over \mu + p} \right],
   \label{SIJ-eq} \\
   & & \dot {\widehat v}_{IJ}
         + H \left[ 1 - {3 \over 2} \left( c_I^2 + c_J^2 \right) \right] \widehat v_{IJ}
         - 3 H \left( c_I^2 - c_J^2 \right)
         \left[ {1 \over 2} \sum_K {\mu_K + p_K \over \mu + p}
         \left( \widehat v_{IK} + \widehat v_{JK} \right)
         + \widehat v \right]
         - {c^2 \over 2a} \left( c_I^2 + c_J^2 \right) S_{IJ}
   \nonumber \\
   & & \qquad
         - {c^2 \over a} \left( c_I^2 - c_J^2 \right)
         \left[ {1 \over 2} \sum_K {\mu_K + p_K \over \mu + p} \left( S_{IK} + S_{JK} \right)
         + {\delta \mu \over \mu + p} \right]
         - {c \over a} e_{IJ}
         - {2 \over 3} {c^2 \over a} {\Delta + 3 K \over a^2} \Pi_{IJ}
         = - {c \over a} I_{IJ}
   \nonumber \\
   & & \qquad
         + {1 \over 2 a} \left( {1 + c_I^2 \over \mu_I + p_I} I_{I0}
         + {1 + c_J^2 \over \mu_J +  p_J} I_{J0} \right) \widehat v_{IJ}
         + {c \over a} \left( {1 + c_I^2 \over \mu_I + p_I} I_{I0}
         - {1 + c_J^2 \over \mu_J +  p_J} I_{J0} \right)
         \left[ {1 \over 2} \sum_K {\mu_K + p_K \over \mu + p}
         \left( \widehat v_{IK} + \widehat v_{JK} \right)
         + \widehat v \right].
   \nonumber \\
   \label{vIJ-eq}
\eea The entropic perturbation, $e$, can be written as (Kodama \& Sasaki 1984)
\bea
   & & e \equiv \delta p - c_s^2 \delta \mu
         \equiv e_{\rm int} + e_{\rm rel}, \quad
         c_s^2 \equiv {\dot p \over \dot \mu}
         = {1 \over \mu + p} \sum_K
         c_K^2 \left( \mu_K + p_K
         + {c \over 3 H a} I_{K0} \right), \quad
         e_{\rm int} \equiv \sum_K e_K, \quad
   \nonumber \\
   & &
         e_{\rm rel} \equiv \sum_K \left( c_K^2 - c_s^2 \right) \delta \mu_K
         = {1 \over 2} \sum_{K,L} {(\mu_K + p_K) (\mu_L + p_L) \over \mu + p}
         \left( c_K^2 - c_L^2 \right) S_{KL}
         - {c \over 3 H a} \sum_K c_K^2 I_{K0} {\delta \mu \over \mu + p}.
\eea
Instead of equations (\ref{eq6-I-linear}) and (\ref{eq7-I-linear}), equations (\ref{SIJ-eq}) and (\ref{vIJ-eq}) together with equations (\ref{eq1-linear})-(\ref{eq7-linear}) provide the complete set of equations.
Various analytic solutions in the system of radiation and dust (baryon plus dark matter) are studied in Kodama \& Sasaki (1986, 1987).

Similar expressions using the relative variables in the case of scalar fields can be found in Hwang \& Noh (2000, 2002).
\end{widetext}

\section{Discussion}

We have been pursuing fully nonlinear and exact formulation of the relativistic cosmological perturbation theory in the context of Friedmann cosmological background. In our previous works we have presented the basic perturbation equations in the presence of a fluid or a minimally coupled scalar field (Hwang \& Noh 2013a, Noh 2014) and applied the equations to various situations: Newtonian limit, Newtonian limit with general relativistic pressure, and post-Newtonian limit (Hwang \& Noh 2013b, 2013c, Noh \& Hwang 2013).

In this work we extend our previous works to include the multiple component fluids and fields. The followings are new results in this work.

1. Fully nonlinear and exact perturbation equations with multiple components of fluids (without anisotropic stress) and minimally coupled scalar fields. We consider the general background curvature and the cosmological constant. The basic sets of the equations are equations (\ref{eq1})-(\ref{eq5}) for Einstein's equations, with equations (\ref{eq6-ADM})-(\ref{eq7-cov}) for fluids, and equation (\ref{EOMs-perturbed}) for fields. Relations among the collective and individual fluid quantities are presented in equation (\ref{fluid-Multi-2}) for fluids, and equation (\ref{fluid-MSFs-exact}) for fields. These are all exact equations readily applicable to fully nonlinear perturbations by the simple Taylor expansion. The fluids formulation in Sec.\ \ref{sec:fluids} also applies to the fields system in Sec.\ \ref{sec:fields} as well as the mixed system with the multiple fluids and fields. The available temporal gauge (slicing or hypersurface) conditions are presented in equations (\ref{temporal-gauges-NL}), (\ref{temporal-gauges-I}) and (\ref{temporal-gauges-I-MSFs}). These slicing conditions can be imposed to any order perturbation with the remaining variables equivalently gauge-invariant to fully nonlinear orders.

2. Newtonian limit in the zero-shear gauge and the uniform-expansion gauge: see equations (\ref{mass-conserv-N-I})-(\ref{U-eq}).

3. Newtonian equations modified by the general relativistic pressures in the zero-shear gauge: see equations (\ref{E-conserv-I-pressure-N}), (\ref{mom-conserv-I-pressure-N}), (\ref{Poisson-eq-pressure}) and (\ref{continuity-p})-(\ref{fluid-p}).

4. Exact and third order perturbation equations in the CDM-comoving gauge. We considered multiple ideal fluids in the presence of one CDM (zero-pressure fluid) component: see equations (\ref{dot-varrho-c-CCG})-(\ref{dot-v-I-CCG}). The equations in the CDM-comoving gauge is convenient to study the nonlinear power spectra (Jeong et al 2011). To the third order, see equations (\ref{dot-varrho-c-third-CCG})-(\ref{dot-phi-p}).

Einstein's equations with a single component fluid is valid even in the presence of the fluids and fields.
However, as the multiple nature of the fluids and fields introduces the anisotropic stress of the collective component, we have extended our previous work on a single ideal fluid or field to include the anisotropic stress in Einstein's equation. We still have assumed vanishing anisotropic stress of the individual fluid component. We find that including the anisotropic stress in the conservation equations is a demanding task, and leave it for future presentation.

In Noh \& Hwang (2013) we have shown that the cosmological first-order post-Newtonian (1PN) equations (modulo anisotropic stress) in Hwang, Noh \& Puetzfeld (2008) can be easily derived from our fully nonlinear cosmological formulation. Our multi-component formulation may allow simple derivation of the 1PN equations in the presence of multiple fluids (modulo anisotropic stress) as well. This subject is also left for a future study.

%
%
\section*{Acknowledgments}
J.H.\ was supported by Basic Science Research Program through the NRF of Korea funded by the Ministry of Science, ICT and Future Planning (No.\ 2013R1A2A2A01068519).
H.N.\ was supported by NRF of Korea funded by the Korean Government (No.\ 2015R1A2A2A01002791).
C.G.P. was supported by Basic Science Research Program through the National Research Foundation (NRF) of Korea funded by the Ministry of Science, ICT and Future Planning (No.\ 2013R1A1A1011107).

%
%


\end{document}